\documentclass[fleqn,usenatbib]{mn2e}
\usepackage{mathptmx}
\usepackage{txfonts}
\usepackage[T1]{fontenc}

\usepackage{graphicx}

\newcommand\pp{{\phantom{$+$}}}
\newcommand\msol{{\cal M_{\odot}}}
\newcommand\teff{{T_{\rm eff}}}
\newcommand\lta{\mathrel{\hbox{\raise 0.6 ex \hbox{$<$}\kern
                   -1.8 ex\lower .5 ex\hbox{$\sim$}}}}
\newcommand\gta{\mathrel{\hbox{\raise 0.6 ex \hbox{$>$}\kern
                   -1.7 ex\lower .5 ex\hbox{$\sim$}}}}

\title[Isochrones for Several Mixtures of the Metals]{Models of Metal-Poor
Stars with Different Initial Abundances of C, N, O, Mg, and Si.~I.~Bolometric
Corrections Derived from New MARCS Synthetic Spectra and Their Implications for
Observed Colour-Magnitude Diagrams} 
\author[D. A. VandenBerg et al.]{
Don A.~VandenBerg,$^{1}$\thanks{E-mail: vandenbe@uvic.ca } 
Bengt Edvardsson,$^{2}$ 
Luca Casagrande,$^{3}$ 
and Jason W.~Ferguson$^{4}$
\\
$^{1}$Department of Physics and Astronomy, University of Victoria, 
P.O. Box 1700, STN CSC, Victoria, BC, Canada V8W 2Y2 \\
$^{2}$Theoretical Astrophysics, Department of Physics \& Astronomy,
Uppsala University, Box 516, SE-751 20 Uppsala, Sweden \\
$^{3}$Research School of Astronomy and Astrophysics, Mount Stromlo Observatory,
The Australian National University, Canberra, A.C.T.~2611, Australia \\
$^{4}$Department of Physics, Wichita State University,
Wichita, KS 67260-0032, U.S.A }

\date{Accepted XXX. Received YYY; in original form ZZZ}
\pubyear{2020}

\begin{document}
\label{firstpage}
\pagerange{\pageref{firstpage}--\pageref{lastpage}}
\maketitle

\begin{abstract}

New, high-resolution MARCS synthetic spectra have been calculated for more than
a dozen mixtures of the metals allowing, in turn, for variations in C:N:O,
[CNO/Fe], and enhanced abundances of C, O, Mg, and Si.  Bolometric corrections
(BCs) for many of the broad-band filters currently in use have been generated
from these spectra.  Due to improved treatments of molecules that involve atoms
of C, N, and O, the BCs for UV and blue passbands, in particular, differ
substantially from those derived from previous MARCS models.  These differences,
and the effects on the BCs of varying the abundances of the metals, are shown in
a number of instructive plots.  Stellar evolutionary grids for $-2.5 \le$ [Fe/H]
$\le -0.5$ have also been computed for the different mixtures.  Isochrones based
on these tracks are intercompared on the theoretical H-R diagram and on a few of
the colour-magnitude diagrams that can be constructed from {\it HST} Wide Field
Camera 3 (WFC3) $F336W, F438W, F606W, F814W, F110W,$ and $F160W$ observations.
For the first application of these models, isochrones have been fitted to WFC3
photometry of the globular cluster NGC\,6496 from the {\it HST} UV Legacy
Survey, with very encouraging results.

\end{abstract}


\section{Introduction}
\label{sec:intro}
 
{\it Hubble Space Telescope} ({\it HST}) Wide Field Camera 3 (WFC3) observations
of globular clusters (GCs) are yielding spectacular colour-magnitude diagrams
(CMDs) that reveal the presence of chemically distinct, multiple stellar
populations (MSPs) within them to an unprecedented degree; see, e.g., the recent
studies of NGC\,2808 (\citealt{mmp15a}), M$\,2$ (\citealt{mmp15b}),
$\omega\,$Cen (\citealt{bma17}), and NGC\,2419 (\citealt{zmm19}).  Indeed, the
{\it HST} UV Legacy Survey (\citealt{pmb15}, \citealt{nlp18})
was designed to use just five filters
($F275W$, $F336W$, $F438W$, $F606W$, and $F814W$) to
identify stars with different abundances of He, C, N, and O.  Complementary
investigations of lower main-sequence (LMS) stars in the IR have also been used
to differentiate between them in terms of their oxygen contents (e.g.,
\citealt{mmb17}, \citealt{mmb19}), and to determine the cluster distances, and
hence ages (e.g., \citealt{dbp15}, \citealt{mfm16}, \citealt{cgk18},
\citealt{sdf18}).

To be sure, the {\it existence} of the MSPs that reside in GCs was already well
established as the result of the enormous amount of spectroscopic work that has
been carried out since the middle of the last century.  Spectra of cluster
giants revealed early on that there are star-to-star variations in the
strengths of the CN bands (\citealt{pop47}, \citealt{osb71}) and the abundances
of Na (\citealt{coh78}, \citealt{pet80}), Mg (\citealt{pil89}, \citealt{she96a}),
and Al (\citealt{ncf81}).  Moreover, the strengths of some
spectral features were found to be either correlated or anti-correlated with the
strengths of other features; e.g., CN-strong stars have larger Na and Al
abundances, but smaller O abundances, than CN-weak stars (\citealt{np85},
\citealt{skp91}, \citealt{dss92}, \citealt{bw92}).  For many years it was
debated (see, e.g., the review by \citealt{kra94}) whether the derived
abundances were primordial in nature or due primarily to nucleosynthesis and
mixing processes within the stars, given that the targets of these studies were
highly evolved giants.

To try to resolve this issue, \citet{hes78} observed lower red-giant branch (RGB)
stars in 47 Tuc, finding that CN variations persisted to at least $M_V \sim +3$.
\citet{hb80} then determined that the variation of CN strengths in several upper
MS stars in the same cluster was comparable to that seen in bright giants, which
argued in support of the possibility that the stars formed out of material with
nonuniform ratios of C:N:O --- unless deep mixing can occur
relatively close to the turnoff (TO).  Many subsequent studies over the next two
decades showed that it is a common property of TO and upper MS stars in GCs that
they exhibit variations in CN, O, Na, Mg, and Al, though the amplitudes of
of such variations vary from cluster to cluster (e.g., \citealt{ccb98},
\citealt{gbb01}, \citealt{rc03}, \citealt{cbs05} \citealt{cbg09b}).  This left
little doubt that the gas out of which the cluster stars formed must have been
chemically inhomogeneous --- although their surface abundances are subsequently
altered to some extent by diffusive processes during MS evolution (e.g.,
\citealt{rmr02}, \citealt{vrm02}, \citealt{dcc17}) and by deep mixing along the
upper RGB (e.g., \citealt{clb82}, \citealt{pil88}, \citealt{dv03}).  More recent
advances include the determination of Mg isotopic ratios (\citealt{she96b},
\citealt[\citealt{yal06}]{ygl03}) and the discovery of the Al--Si correlation
(\citealt{ygn05}), which are both indicative of of $p$-capture nucleosynthesis
at very high temperatures. 
%Variations in
%C$+$N$+$O in $\omega\,$Cen (\citealt{          and NGC\,1851 (\citealt{dgn15})
%\footnote{At the present time,
%only supermassive stars {\citealt{dh14} appear to have the necessary temperature
%structures to explain the isotopic abundances and the various abundance
%correlations and anticorrelations {see \citealt{
%provide strong contraints on the temperature of the nucleosynthesis site,
%isotopic ratios in NGC\,6752 by \citet{yab06}, the evaluation of O--Na and
%Mg--Al anticorrelations in 17 GCs by \citet{cbg09b}, and  C+N+O is not constant
%is some systems ...
%
%, Al (\citealt{nor81}), and O
%(\citealt{lwo86}).  Notable examples of the
%impressive progress that has been made on this
%front include the analyses of CN strengths, and the inferred (large) ranges in N
%abundances by \citet{cbs05}, \citet{aaa10}, and \citet{ggg10}, the determination
%of Mg isotopic ratios in NGC\,6752 by \citet{yad06}, the evaluation of the
%O--Na and Mg--Al anticorrelations in 17 GCs by \citet{cbg09b}, the separation of
%stars in $\omega\,$Cen into five major groups based on distinctive chemical
%abundance differences by \citet[also see \citealt{jp12} for similar work on
%M\,13]{jp10}, and the detection of exceptionally large Mg abundance variations
%in NGC\,2419 \citet{ck12}.  The advances that has been made over the years
%are well documented in the reviews by, e.g., \citet{kkk94}, \citet{ggg1}, and
%\citet{ggg}.

It is likely to be several years before reliable quantitative measurements of
light element abundances will be obtained from photometry because bolometric
corrections (BCs), at short wavelengths in particular, require that the 
synthetic spectra from which they are calculated take into account all of the
important sources of opacity due to atomic lines and molecular bands.  This is
a huge, ongoing challenge, especially in the UV.  Contributing to this problem 
are the uncertainties associated with the stellar $\teff$\ scale, which appear
to be at the level of $\sim 70$--100~K for upper MS and TO stars
(\citealt{crm10}), but undoubtedly much higher than this for the coolest dwarfs
and giants.  Also of considerable importance are the uncertainties connected
with the microturbulence, which is used as a free parameter in one-dimensional
model atmospheres to broaden spectral lines in order to take into account
small-scale motions generated by atmospheric convection.  Because the
microturbulence has the effect of redistributing the flux in spectral regions
that are crowded with lines, it can affect BCs in some passbands by several
hundredths to a few tenths of a magnitude, depending on the temperature and
metallicity (\citealt[hereafter CV14; see their Figs.~3 and 4]{cv14}).  (The
fluxes computed from 3D model atmospheres are not subject to this limitation as
they are derived from the modeling of convective motions and turbulent flows;
see, e.g., \citealt{bcl18}, \citealt{ccc18}.)  The superb CMDs from the UV
Legacy Survey (\citealt{nlp18}) will clearly be an invaluable resource for the testing
and improvement of synthetic colour--$\teff$ relations that are used by
isochrones to interpret these observations.

A possible indication of the deficiencies of current stellar models and
synthetic spectra is the finding by \citet{mmr18} that the helium abundances
derived from so-called ``chromosome maps" (\citealt{mmp15a}), which use a
specific combination of UV and optical colour indices to provide a clear
separation of stars with different chemical compositions, do not agree with the
abundance variations that have been inferred from theoetical sumulations of
horizontal-branch (HB) populations in GCs.\footnote{This problem does raise any
doubts about the very high He abundance that has been deduced for the blue MS
stars in $\omega\,$Cen (\citealt{pvb05}), which is arguably the most important,
and the most surprising, discovery in the stellar populations research area in
recent years.  Isochrones appear to be able to fit the CMD locations of these
stars only if they assume $Y\sim 0.4$ (see, e.g., \citealt{kbc12},
\citealt{hvn12}).}  For instance, the latter deduced that $\delta\,Y \gta 0.10$
in M\,3 from the length of the sequence of stars in the chromosome map that they
ascribe to first-gneration stars, whereas they obtained $\delta\,Y_{\rm max} =
0.004\pm 0.011$ for NGC\,6362.  This is in stark contrast with the results of
\citet{dvk17}, who found a very good match between the predicted and observed
HB of M\,3 if $\delta\,Y \sim 0.01$ and the models assume a moderate amount of
mass loss along the giant branch.  [A follow-up study of M\,3 by \citet{tdc19}
also concluded that the large spread in $Y$ derived from the chromosome map was
incompatible with the cluster HB morphology, the period and luminosity
distribution of its RR Lyrae, and the colour distribution of its MS stars.]  In
the case of NGC\,6262, the difference in luminosity between the non-variable HB
stars on either side of the instability strip, as well as simulations of the
entire HB population, imply $\delta\,Y \sim 0.03$ (\citealt{vd18}).  It would
appear that our understanding of chromosome maps, or of GCs, is lacking in some
fundamental way. 

\begin{table*}
\centering
\caption{The Adopted Chemical Abundances and the Associated BC Tables} 
 \label{tab:t1}
\smallskip
\begin{tabular}{lcccccccc}
%\tablewidth{600pt}
%\tablewidth{0pt}
\hline
\hline
\noalign{\smallskip}
 Names of & & & & & & & & \\
 BC Tables$^{a}$ & He & C & N & O & [CNO/Fe] 
   & Mg & Si & $v_T^{b}$ \\
\noalign{\smallskip}
\hline
\noalign{\smallskip}
%\startdata
%\noalign{\vskip 1pt}
%\multispan11 {\hfil [$\alpha$/Fe varies with [Fe/H] \hfil} \\
%\noalign{\vskip 3pt}
 \bf{a4s08}  & 10.93 & \pp8.39 & \pp7.78 & \pp9.06 & $+0.28$ &
   \pp7.93 & \pp7.91 & 2.0 \\
 a4s21  & 11.00 & \pp8.43 & \pp7.83 & \pp9.09 & $+0.28$ &
   \pp8.00 & \pp7.91 & f(g) \\
 a4CN   & \bf{--} & $\downarrow\,0.3$ & $\uparrow\,0.50$ & \bf{--} &
  $+0.28$ & \bf{--} & \bf{--} & f(g) \\
 a4CNN  & \bf{--} & $\downarrow\,0.3$ & $\uparrow\,1.13$ & \bf{--} & $+0.44$ &
   \bf{--} & \bf{--} & f(g) \\
 a4ON   & \bf{--} & $\downarrow\,0.8$ & $\uparrow\,1.30$ & $\downarrow\,0.8$ &
   $+0.28$ & \bf{--} & \bf{--} & f(g) \\
 a4ONN   & \bf{--} & $\downarrow\,0.8$ & $\uparrow\,1.48$ & $\downarrow\,0.8$ &
   $+0.44$ & \bf{--} & \bf{--} & f(g) \\
 a4xC$\_$p4   & \bf{--} & $\uparrow\,0.4$ & \bf{--} & \bf{--} & $+0.38$ &
   \bf{--} & \bf{--} & f(g) \\
 a4xCO   & \bf{--} & $\uparrow\,0.7$ & \bf{--} & $\uparrow\,0.2$ & $+0.61$ &
   \bf{--} & \bf{--} & f(g) \\
 a4xO$\_$p2  & \bf{--} & \bf{--} & \bf{--} & $\uparrow\,0.2$ & $+0.44$ &
   \bf{--} & \bf{--} & f(g) \\ 
 a4xO$\_$p4  & \bf{--} & \bf{--} & \bf{--} & $\uparrow\,0.4$ & $+0.62$ &
   \bf{--} & \bf{--} & f(g) \\ 
 a4Mg$\_$p2 & \bf{--} & \bf{--} & \bf{--} & \bf{--} & $+0.28$  &
   $\uparrow\,0.2$ & \bf{--} & f(g) \\
 a4Si$\_$p2 & \bf{--} & \bf{--} & \bf{--} & \bf{--} & $+0.28$ & \bf{--} &
   $\uparrow\,0.2$ & f(g) \\
\noalign{\smallskip}
 \bf{s08std}  & 10.93 & \pp8.39 & \pp7.78 & \pp8.66 & $0.00$--$0.28^{c}$ &
   \pp7.53 & \pp7.51 & 2.0 \\
 s08vt1 & \bf{--} & \bf{--} & \bf{--} & \bf{--} & \bf{--} & \bf{--} &
   \bf{--} & 1.0 \\
 s08vt5 & \bf{--} & \bf{--} & \bf{--} & \bf{--} & \bf{--} & \bf{--} &
   \bf{--} & 5.0 \\
 s21std & \bf{--} & \bf{--} & \bf{--} & \bf{--} & \bf{--} & \bf{--} &
   \bf{--} & f(g) \\
 s21Y30 & 11.04 & \bf{--} & \bf{--} & \bf{--} & \bf{--} & \bf{--} &
   \bf{--} & f(g) \\
\noalign{\smallskip}
\hline
\noalign{\smallskip}
\end{tabular}
%\noalign{\vskip 2pt}
%\enddata
\begin{minipage}{1\textwidth}
$^{a}$~Boldface font identifies reference models (see the text); the others
involve changes to the abundances of one or more of the metals, as
tabulated, or to $v_T$.  Only the ``a4" BCs are used to transform isochrones
to various CMDs; the others are employed by codes that are provided (see the
Data Availability section) to evaluate, and to present in tabular form, the
effects on BCs of varying the microturbulence or the assumed helium abundance. \\ 
$^{b}$~f(g) implies that the microturbulent velocity, $v_T$, varies with gravity
such that $v_T = 1.0$ km/s if $\log\,g \ge 4.0$ or 2.0 km/s if $\log\,g \le
3.0$. \\
$^{c}$~The specified range in [CNO/Fe] corresponds to an increase in
[O/Fe] from 0.0 at [Fe/H] $= 0.0$ to 0.4 at [Fe/H] $= -1.0$ (see the text).\\
\phantom{~~~~~~~~~~~~~~~~~~~~~~~~~}
\end{minipage}
\end{table*}

During the past several decades, many studies have examined the effects on
stellar models of varying the assumed metal abundances.  For instance, variations
in O were considered by \citet{rc85} and \citet{van92}, while differences in the
abundances of the $\alpha$-elements as a group were studied by \citet{scs93},
\citet{vsr00}, \citet{pcs06}, and \citet{dcf07}, among others.
\citet[also see \citealt{bnf16}]{vbd12} computed isochrones for metal-deficient
stars in which 10 of the most abundant metals from C to Ti were enhanced by 0.4
dex, in turn, at constant [Fe/H].  Colour--$\teff$\ relations for different
choices of [$\alpha$/Fe] have been generated by CV14 (also see \citealt{csc04})
for large ranges in $\log\,g$, $\log\teff$, and [Fe/H], based on MARCS model
atmospheres and synthetic spectra (\citealt{gee08}).  Investigations of the
consequences of C-N-O-Na-Mg-Al correlations and anticorrelations for isochrones
and their transformations to various CMDs were undertaken by \citet{swf06},
\citet{pcs09}, \citet{ssw11}, and \citet{cmp13}.  This brief mention of just a
few of the papers that have dealt with stellar abundances from a theoretical
perspective is sufficient to show that the most basic questions concerning the
impact of abundance variations have already been addressed.  It seems likely
that further advancements will mainly occur as a result of refinements to model
atmospheres, synthetic spectra and the associated BCs, and the $\teff$\ scale of
stellar evolutionary computations.

Because the CV14 transformations, coupled with Victoria-Regina isochrones
(\citealt{vbf14}) have had considerable success reproducing optical and near-IR
CMDs of GCs (see, e.g., \citealt[hereafter VBLC13]{vbl13}; \citealt{cgk16}), we
decided to compute new synthetic spectra that take various abundance variations
into account, to calculate BCs from those spectra for a large fraction of the
broad-band filters currently in use, including the {\it HST} WFC3 filters, and
to generate new isochrones for the assumed abundances.  It turns out that the
improvements which have been made by the Uppsala group to the MARCS spectral
synthesis code since the study by \citet{gee08} --- primarily to the treatment
of molecules involving C, N, and O --- have very significant consequences for
the BCs that are calculated for UV filters but rather little impact on the those
at optical and longer wavelengths.  However, as noted by \citet{edv08}, MARCS 
model fluxes are less reliable in the blue spectral region due to the importance
of, and uncertainties associated with, the continuous non-hydrogenic opacities 
at shorter wavelengths.

The main goal of this series of papers is to examine how well isochrones
that employ the improved BCs are able to reproduce observed GC CMDs that
involve colours ranging from $m_{F336W}-m_{F438W}$ to $m_{F110W}-m_{F160W}$.
Unfortunately, the synthetic spectra do not extend far enough into the UV to
permit the calculation of BCs for the $F275W$ filter, which was selected for
the {\it HST} UV Legacy Survey to distinguish between stars with different O
abundances because this passband contains an OH band; see \citealt{pmb15}.
In this paper, we describe and discuss the assumed heavy-element mixtures and
their consequences for both the BCs and stellar evolutionary models.  
Comparisons of our isochrones with WFC3 observations of GCs that span a wide
range in [Fe/H] are presented in Paper II.  
%This will be followed by an
%investigation, in Paper III, of the complex stellar populations that reside
%in $\omega\,$Cen.

\section{The Calculation of Bolometric Corrections}
\label{sec:calBC}

\subsection{The Assumed Chemical Abundances}
\label{subsec:mix}

Table~\ref{tab:t1} lists the $\log\,N_i$ number abundances, on the scale
$\log\,N_{\rm H} = 12.0$, of just those elements for which abundance variations 
have been considered in this investigation.  The two reference models (in
boldface font) assume the solar abundances reported by \citet{gas07}, which were
adopted in the generation of large grids of MARCS model atmospheres and
synthetic spectra by \citet{gee08}.  Whereas {\tt a4s08} takes into account a
0.4 dex enhancement of all of the $\alpha$ elements (O, Mg, and Si, as well as
Ne, S, Ca, and Ti, even though they are not explicitly included in the table),  
{\tt s08std} assumes that [$\alpha$/Fe] $= +0.4$ at [Fe/H] $\le -1.0$, with a
linear decrease from $0.4$ at [Fe/H] $= -1.0$ to 0.0 at [Fe/H] $= 0.0$ (a
pattern that is roughly characteristic of local Galactic disk and halo stars;
e.g., \citealt{eag93}).  In the case of the {\tt a4s08} models, the abundances
of all of the metals at a given metallicity can be obtained simply by adding the
[Fe/H] value to the tabulated $\log\,N_i$ values.  For the {\tt s08std} models,
the abundances of the $\alpha$ elements must first be increased by the relevant
value of [$\alpha$/Fe] (e.g., 0.3 dex if considering stars with [Fe/H] $= -0.75$)
before the abundances of all of the metals are scaled to the metallicity of
interest ([Fe/H] $= -0.75$ in our example).

In order that the computed grids of evolutionary tracks and isochrones be 
consistent with those reported by \citet{vbf14}, which allow for variations in
[$\alpha$/Fe] over wide ranges in [Fe/H] and $Y$, we decided to adopt the
elemental abundances given by \citet{ags09} as the base mixture, instead of
\citet{gas07}.  Thus, in Table~\ref{tab:t1}, the name {\tt a4s21} has been
given to the mixture of the metals reported by Asplund et al.~with a 0.4 dex
enhancement of the $\alpha$ elements.  Note that ``a4" implies that
``[$\alpha$/Fe] $= +0.4$" in the base mixture and that the numbers ``08" or
``21" in the names identify, in turn, the 2008 or 2021 MARCS models.  To
examine the impact of varying C, N, O, Mg, and Si, the {\tt a4s21} abundances
have been modified in a number of different ways by amounts that are described
by upward- or downward-pointing arrows followed by the adopted values of
$\delta\,\log\,N_i$. 

The assigned names of the different variations were chosen to make it easy to
remember them.  For instance, for typical scaled-solar abundances of C and N,
the CN abundance reaches a maximum value due to CN-cycling when C has been
reduced by $\sim 0.3$ dex (see, e.g., \citealt[his Fig.~8]{smi87}), which then
requires that N be enhanced by $\sim 0.5$ dex if C$+$N $= constant$.  Just as
{\tt a4CN} is a reasonable label for this case, {\tt a4CNN} is a suitable name
for a mixture that assumes the same depletion of C but a higher N abundance.  
Similarly, predictions of ON-cycling by, e.g., \citet{dw04} are able to explain
the very low C and O abundances ([C/Fe] $\sim -0.8$, [O/Fe] $\sim -0.4$) that
have been derived spectroscopically in some GC stars (see, e.g., \citealt{ssb96},
\citealt{cbs02}).  Accordingly, a mixture of the metals in which the {\tt a4s21}
abundances of C and O have been depleted by 0.8 dex, and N increased by an
amount such that C$+$N$+$O $= constant$, has been given the name {\tt a4ON},
while {\tt a4ONN} refers to a mixture having the same depletions of C and O but
a higher N abundance --- to be consistent with a higher value of [CNO/Fe].
Spectroscopic work over the years has established that values of [N/Fe] $\sim
+1.5$ are typically found in GCs; see \citet{bcs04}, \citet{sbh05},
\citet{cbs05}.

As $\alpha$-enhanced Population II stars appear to have [C/Fe] $> 0$
(specifically, [C/Fe] $\approx 0.2$; see
\citealt{ncc14}), we opted to consider two cases with increased C abundances;
specifically, {\tt a4xC\_p4}, which assumes a 0.4 dex enhancement, and
{\tt a4xCO}, which considers higher abundances of both C and O by 0.7 dex
and 0.2 dex, respectively.  When the abundances of single elements are altered,
we have adopted the convention wherein the symbol for the element is followed by
``\_{\tt p}", where ``{\tt p}" represents ``$+$" and the number after the
``{\tt p}" gives the enhancement in tenths of a dex.  Granted, these abundances
of carbon are much higher than those derived in current observational studies of
GCs (see, e.g., \citealt{cgl05}), though high C is suggested by anomalous
populations of CO-strong stars in some systems (notably $\omega\,$Cen; see
\citealt[and references therein]{ndc95}).  (The determination of whether or not
[C/Fe] varies with [Fe/H] depends quite sensitively on the adopted $\teff$ scale
and on the importance of non-LTE effects, as shown in the study of metal-poor
field stars by \citealt{fna09}.)  In addition, Paper II provides some
tantalizing evidence that GCs may contain sub-populations of stars with unusually
high C enhancements; consequently, there is ample justification for studying the
{\tt a4xC\_p4} and {\tt a4xCO} cases.  The remaining mixtures that we have  
%Given the above remarks, it should be
% Regardless, since it is the intention of
%D.A.V.~to provide grids of stellar models in which [C/Fe], [O/Fe], [Mg/Fe], and
%[Si/Fe] are treated, in turn, as free parameters, the models produced for this
%study will be incorporated into the eventual grids that can be interpolated for
%intermediate abundance enhancements.  
%Model atmospheres, synthetic spectra, and tables of BCs have also
considered allow for 0.2 and 0.4 dex enhancements of oxygen
({\tt a4xO\_p2} and {\tt a4xO\_p4}, respectively), as well as enhanced Mg by 0.2
dex ({\tt a4Mg\_p2}) and increased Si abundances by 0.2 dex
({\tt a4Si\_p2}).\footnote{During the course of this
investigation, we found that the effects on BCs of an 
enhanced Mg abundance by 0.2 dex were significantly larger than those obtained
by increasing the abundances of all of the $\alpha$ elements by 0.2 dex.
B.E.~and his colleague at Uppsala Observatory, K. Eriksson, studied this
difficulty and found that it occurred because of an assumption that was made
when considering modest abundance enhancements of single elements.  Normally,
synthetic spectra are based on fully consistent model atmospheres.  However, in
the case of Mg, the model atmosphere structures were initially assumed to be the
same with, or without, the 0.2 dex enhancement in its abundance; i.e., the
abundance difference was treated only in the calculation of the synthetic
spectra.  When B.E.~and K.E.~computed model atmospheres for enhanced Mg and
based the synthetic spectra on these atmospheres, the resultant BCs were much
more consistent with those obtained when the abundances of all of the $\alpha$
elements as a group were increased by the same amount.  This is potentially a
very important finding that should be thoroughly studied.  The BCs at short
wavelengths, in particular, where there is little flux in the case of cool
giants and MS stars, are unlikely to be very trustworthy unless they are based
on fully consistent model atmospheres and synthetic spectra.  Indeed, it is
possible that similar consistency may be necessary when determining stellar
abundances from fits to observed spectral features; at least, this is a concern
that warrants careful consideration.  As a result of this discovery, all of our
synthetic spectra are based on fully consistent model atmospheres.}

Although the BCs for the ``{\tt a4}" mixtures have been newly computed 
specifically for this investigation, using the same formalism, zero-points,
and filter transmission curves of CV14, those for the {\tt a4s08} and {\tt
s08std} cases have been derived from the tables given by CV14 using their
software.  These BCs assume a microturbulent velocity, $v_T$, of 2.0 km/s
whereas the default assumption for MARCS models, including the current ones in
the ``{\tt a4}" series, is to adopt 1.0 km/s if $\log\,g \ge 4.0$ and 2.0 km/s
if $\log\,g \le 3.0$.   This implications of this difference for computed BCs
are discussed in \S~\ref{sec:bcs}.

\begin{figure}
\includegraphics[width=\columnwidth]{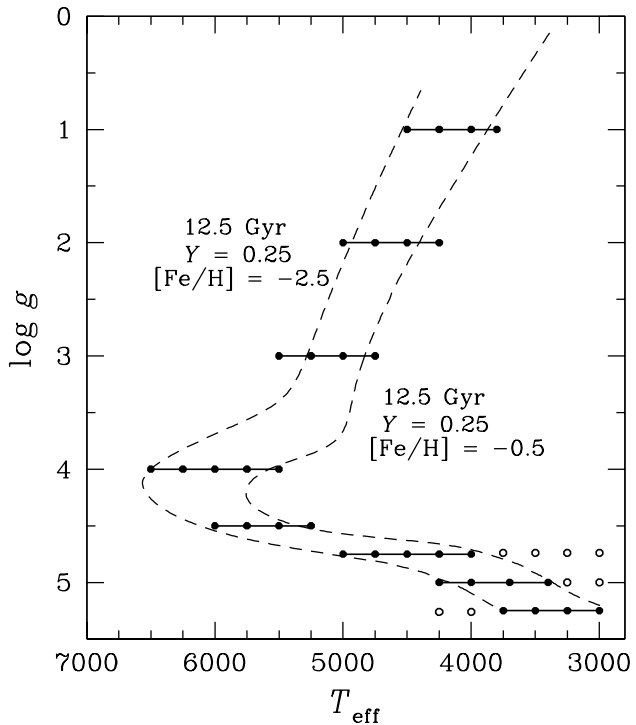}
\caption{A $\teff$, $\log\,g$ diagram that gives the effective temperatures and
gravities (the filled circles) for which MARCS model atmospheres and synthetic
spectra have been computed.  The horizontal solid lines specify the ranges in
$\teff$\ at the adopted $\log\,g$\ values; they match quite well the differences
in $\teff$ predicted by Victoria-Regina isochrones (\citealt{vbf14}) for the
indicated ages and chemical compositions (the dashed loci).  Additional model
atmospheres for use as atmospheric boundary conditions in the stellar models of
low-mass stars were computed for the $\teff, \log\,g$ values that are
represented by small open circles.}
\label{fig:f1}
\end{figure}

\begin{figure}
\includegraphics[width=\columnwidth]{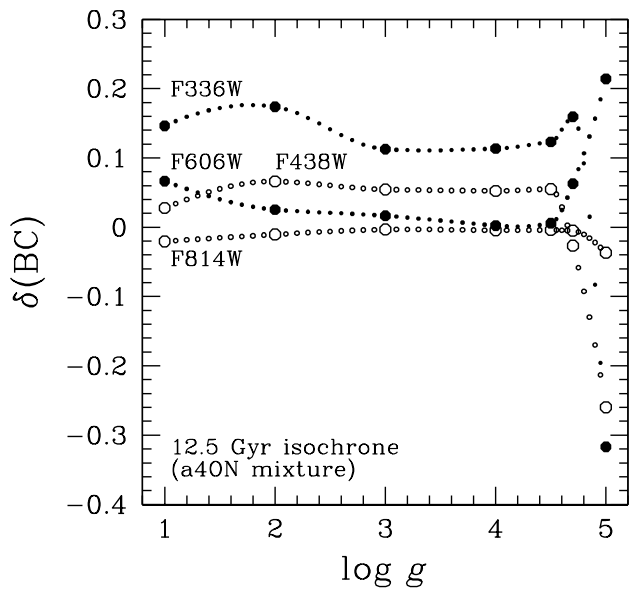}
\caption{The differences between the BCs for the {\tt a4ON} mix and those given
by CV14 for the ({\tt a4s08}) reference models at the $\teff$ values predicted
by a 12.5 Gyr isochrone for [Fe/H] $= -0.5$ and the {\tt a4ON} mix at $\log\,g =
1.0$, 2.0, 3.0, 4.0, 4.5, 4.7, and 5.0 (the large filled and open circles).  The
smaller circles were obtained by interpolating between the larger ones using
Akima splines (see the text).}
\label{fig:f2}
\end{figure}

The ``{\tt a4}" computations also assume $log N_{\rm He} = 11.0$, or $Y \approx
0.28$.  Unlike the metals, helium contributes very little to the
opacities in the surface layers of lower-mass stars, though higher $Y$ does make
the gas slightly more transparent by diluting the electron density and the
metals.  Its main effect is to increase the mean molecular weight, which
compresses the atmosphere somewhat, thereby mimicking an atmosphere with a 
higher gravity.  As reported by \citet{gcb07}, and confirmed by a number of
computations that we have carried out for $Y=0.25$ ({\tt s21std}) and $Y=0.30$
({\tt s21Y30}), moderate He abundance variations do not affect the BCs for
broad-band filters by more than a few to several thousandths of a magnitude;
consequently, it is not necessary to generate BC tables for different values of
$Y$, especially for an exploratory study such as ours.

%{\it Note that only the BCs for the ``a4" series of models can be used to
%transform isochrones to various CMDs; the others are employed by codes that are
%made available to interested users (see the Data Availability section) to
%evaluate, and to present in tabular form, the effects on BCs of varying the
%microturbulent velocity or the assumed helium abundance.}
%Because synthetic spectra are affected only slightly by the
%assumption of different He abundances, if all other parameters are held
%constant, the BCs derived from them will also be nearly identical.  This can
%be verified by comparing the BCs for the {\tt s21std} and {\tt s21Y30}
%mixtures (see Table~\ref{tab:t1}), which differ only insofar as they assume
%a different He abundance by the equivalent of $\delta\,Y \approx
%0.05$.  At the same values of $\teff$, $\log\,g$, and
%[Fe/H], the BCs for broad-band filters differ by just a few thousandths of a
%magnitude (also see \citealt{gcb07}), rising to $\sim 0.015$ mag in the case
%of the $F336W$ filter at $\log\,g \le 1.0$ or $\ge 5.0$, with much smaller
%differences at intermediate gravities. 
%
%There is clearly no need to generate BC tables for different values of $Y$,
%especially for an exploratory study such as ours.  We therefore chose to
%calculate all of the BC tables for the ``{\tt a4}" series assuming $\log N_{\rm
%He} = 11.00$ ($Y \approx 0.28$) and to use these tables to transform stellar
%models for this or any other He abundance to CMDs of interest.

\subsection{The Adopted Parameter Variations}
\label{subsec:param}

The calculation of high-resolution spectra for wide ranges in $\log\,g$,
$\teff$, and [Fe/H] becomes an especially computationally demanding project if
many different mixtures of the heavy elements are also considered.  In order to
make the BC computations more manageable, they have been limited to the
$\log\,g$ and $\teff$ values that are indicated by the filled circles in
Figure~\ref{fig:f1}.  As shown in this figure, the adopted temperatures were
determinated from a consideration of the differences in $\teff$, at the selected
gravities, between isochrones relevant to old stellar populations with [Fe/H]
$= -2.5$ and $-0.5$.  However, even with these restrictions, it was necessary
to generate over 1100 model atmospheres and synthetic spectra ---  for 34
$\teff, \log\,g$\ values at each of three metallicities ([Fe/H] $= -2.5, -1.5$,
and $-0.5$) and each of the 11 ``{\tt a4}" mixtures that are listed in
Table~\ref{tab:t1}).  As the spectra were calculated for $R > 225,000$, where
$R$ is the ratio of the wavelength to the step in wavelength at which the fluxes
are evaluated, spectral features are very well resolved.

%Despite the relatively coarse grid spacing of the synthetic spectra and the BCs
%derived from them, especially with regard to $\log\,g$, we believe that it is
%sufficient to accomplish the transposition of stellar models from the
%theoretical H--R diagram to various CMDs without introducing interpolation
%errors that are large enough to significantly affect the findings of this
%exploratory study.  
In general, the effects on BCs {\it for broad-band filters}
of varying the abundances of a small subset of the metals are not very large
at $\log\,g \lta 4.7$.  Moreover, the variations of the differences in the BCs
that are obtained with, and without, the changes to the mixtures of the metals
considered here tend to be smooth, well-behaved functions of $\log\,g$ that can
be interpolated to quite high accuracy.  Even at $\log\,g > 4.7$, where such
differences tend to increase quite rapidly, the loci connecting the $\delta$(BC)
values at $\log\,g = 4.5$ and $5.0$ appear to be quite well constrained by the
results at $\log\,g = 4.7$.  

These claims are supported by Figure~\ref{fig:f2},
which plots, for four of the WFC3 filters, the differences in the BCs for the
{\tt a4ON} mix and those given by CV14 for the relevant reference mix
({\tt a4s08}) at the fiducial $\log\,g$ values.  To be more specific, both sets
of BCs were derived for the temperatures and gravities along a 12.5 Gyr
isochrone for [Fe/H] $= -0.5$ and the {\tt a4ON} mixture, for which the
$\delta$(BC) variations are comparable to, or larger, than those predicted for
most of the mixtures listed in Table~\ref{tab:t1}. 
%The differences in the
%BCs are also reduced if lower [Fe/H] values are considered.
Interpolations via Akima splines (\citealt{aki70}), which were fitted to the
large filled and open circles at the grid values of $\log\,g$, yielded the 
smaller circles at intermediate gravities.  Even when the $\delta$(BC) values change
by a relatively large amount --- as in the BCs for the $F336W$ filter between
between $\log\,g = 4.7$ and 5.0 (see Fig.~\ref{fig:f2}) --- the interpolated
points still look very reasonable.

\begin{figure*}
\includegraphics[width=\textwidth]{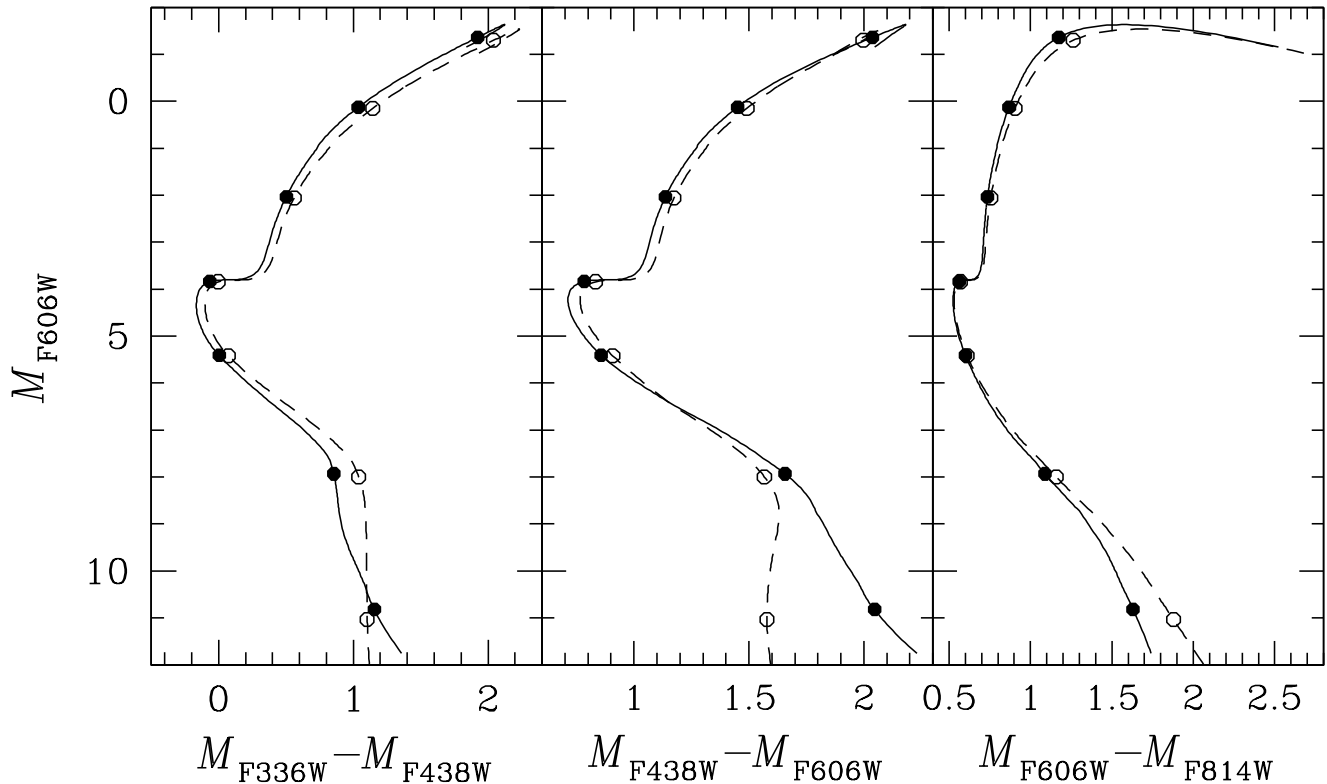}
\caption{Transformation of a 12.5 Gyr isochrone for [Fe/H] $= -0.5$ and the
{\tt a4ON} mixture to three CMDs using CV14 BCs for the standard {\tt a4s08}
mix (dashed curves) and the adjusted BCs that are derived from the $\delta$(BC)
offsets shown in Fig.~\ref{fig:f2} (solid curves).  The large open and filled
circles identify the points along the isochrones at the fiducial $\log\,g$
values (1.0, 2.0, $\ldots$, 5.0).}  
\label{fig:f3}
\end{figure*}

% though one can certainly question whether
%the sudden increase actually occurs right at $\log\,g = 4.5$ or at a slightly
%larger or smaller value.  However, such a large change in $\delta$(BC) over a
%small range in $\log\,g$ is quite rare; i.e., it is not seen in similar plots
%for most of the chemical abundance choices that are listed in Table~\ref{tab:t1}.

To predict the consequences of the different metal abundance variations on a
given CMD, we have therefore opted to use the following procedure.  First,
isochrones for the revised mixture (e.g., {\tt a4ON}) are transposed to the
selected CMDs using the CV14 BCs for the {\tt a4s08} reference mix, which cover
much wider ranges in [Fe/H], $\teff$, and $\log\,g$, at a significantly higher
resolution, than the BC tables that were generated for this study.  Second,
the differences between the CV14 BCs at $\log\,g = 1.0, 2.0, \ldots, 5.0$,
at the isochrone $\teff$\ values, and those for the revised mix are evaluated,
and Akima splines are fitted to the resultant $\delta$(BC) values (as in
Fig.~\ref{fig:f2}).  The last step is simply to use the Akima splines to correct
the CV14 BCs at each of the isochrone points with gravities in the range $1.0
\le\log\,g \le 5.0$.  Extrapolations to lower or higher values of $\log\,g$ are
carried out, as necessary, using quadratic equations that describe the
variations of $\delta$(BC) with gravity at $\log\,g = 1.2$, 1.1, and 1.0 or at
$\log\,g = 4.9$, 4.95, and 5.0, using the interpolated $\delta$(BC) at other
than the grid values of $\log\,g$. 

Figure~\ref{fig:f3} illustrates the results that are obtained for the same
12.5 Gyr isochrone for the {\tt a4ON} mixture that was considered in the
previous figure.  The dashed loci in each panel represents the isochrone when
it is transposed to three of the CMDs that are considered in our analyses of GC
observations in Paper II using CV14 BCs for the $F336W$, $F438W$, $F606W$, and
$F814W$ filters.  The solid curves are obtained if the differences between the
CV14 and the {\tt a4ON} BCs, as derived from Akima spline interpolations
(Fig.~\ref{fig:f2}) and quadratic extrapolations, are applied to the CV14
transformations.  For the most part, the offsets are quite small and they
reflect the behavior of the $\delta$(BC) variations with $\log\,g$\ in the
previous figure; e.g., the cross-over of the $\delta$(BC) values for $F438W$
and $F606W$\ filters translates to a cross-over of the $M_{F438W}-M_{F606W}$
colours at $\log\,g \sim 1.4$ and $\sim 4.6$.\footnote{Although BCs were
%Even when the corrections to the
%BCs and the resultant isochrones are quite large, as in the middle panel at
%$M_{F606W} \gta 9$, the morphological variations are approximately in line with
%expectations.  
generated for $\log\,g = 5.3$ (see Fig.~\ref{fig:f1}), it turns out that stellar
models for Pop.~II stars with such gravities are very close to the H-burning
limit, where the equation of state that we use for low-mass stars, 
FreeEOS (http://freeeos.sourceforge.net) is known to have convergence
difficulties.  Indeed, for some of the metal abundance mixtures, we were unable
to obtain converged MS models with sufficiently low masses ($\lta 0.1 \msol$) in
order to reach $\log\,g \ge 5.3$.  However, it is clear from Fig.~\ref{fig:f3}
that the extensions of our isochrones to $\log\,g = 5.0$ already extend to
$M_{F606W} \sim 11.5$, which is more than adequate for the comparisons with
observations that are presented in Paper II.}  
%Before discussing the evolutionary
%computations, it is worthwhile to examine in some detail how the BCs that we
%have produced at fixed values of $\teff$, $\log\,g$, and [Fe/H] vary with the
%assumed mixtures of the metals.  This is the subject of the next section.

\section{The Effects of Chemical Abundance Variations on Bolometric
Corrections}
\label{sec:bcs}

\begin{figure*}
\includegraphics[width=\textwidth]{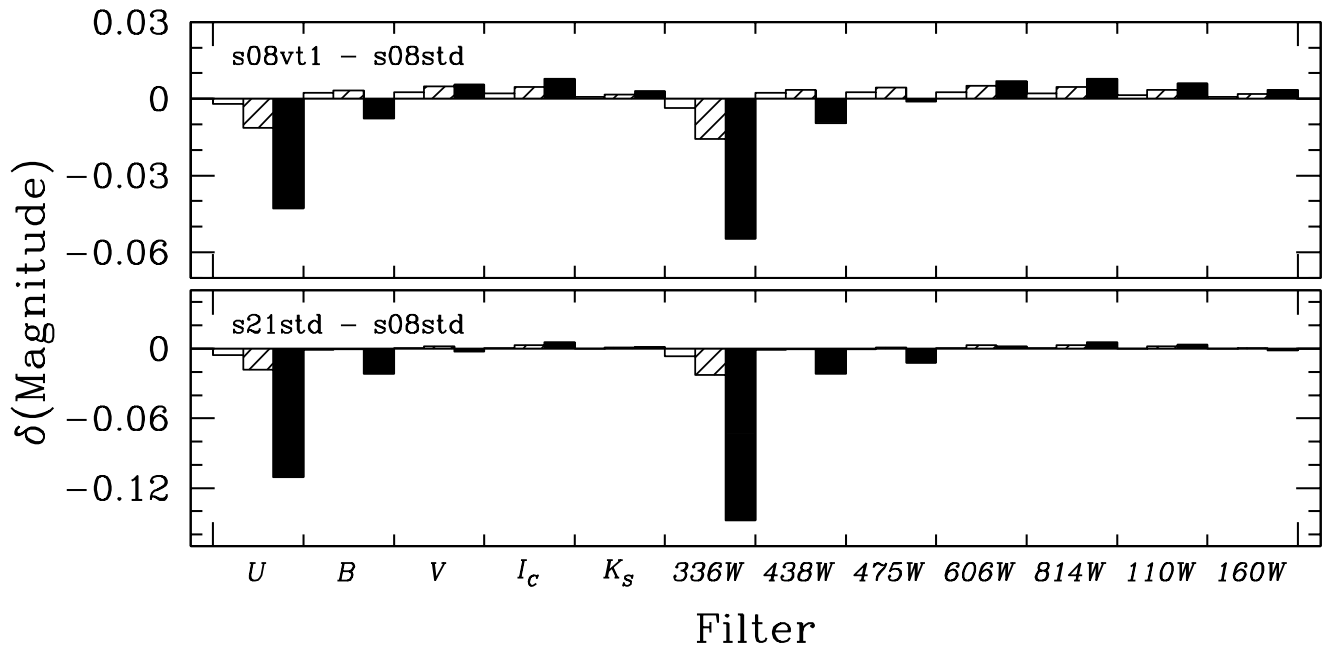}
\caption{{\it Top Panel:} Histogram showing the predicted magnitude change
in passbands ranging from $U$ to $F160W$ if the MARCS models assume a
micro-turbulent velocity, $v_t$, of 1 km/s ({\tt s08vt1} in Table~\ref{tab:t1})
instead of 2 km/s ({\tt s08std}).  The open, hatched, and filled vertical bars
associated with each filter represent the predictions for [Fe/H] $= -2.5,
-1.5$, and $-0.5$, respectively.  The models assume $\log$\,$g = 4.0$ and the
$\teff$s from 12.5 Gyr isochrones at this gravity and the relevant metallicity;
specifically $\teff = 6499, 6162,$ and 5576~K, in turn.  (The ``$F$" in
``$F336W$" and in each of the other WFC3 filter names has been omitted for the
sake of clarity.)  {\it Bottom panel:} Similar to the top panel, except that the
differences between the BCs calculated for the {\tt s21std} and {\tt s08std}
models are shown.  Note that a difference in the assumed microturbulent velocity
contributes to the differences arising from improvements to the MARCS codes.}
\label{fig:f4}
\end{figure*}

To illustrate some of the properties of the different BC transformations, we 
decided to plot histograms of the differences in the BCs that have been derived
for several of the cases in Table~\ref{tab:t1} at common values of $\log$\,$g$,
$\teff$, [Fe/H], and [$\alpha$/Fe]. In fact, we have adopted $\delta$(magnitude)
$= - \delta$(BC) for the ordinate in order to better relate the results to
observed stars.  In addition, we opted to consider the Johnson-Cousins-2MASS
$UBVI_CK_S$ passbands along with seven of the WFC3 filters, ranging from $F336W$
in the UV to $F160W$ in the IR, in order to sample the synthetic spectra over
nearly the full range in wavelength for which they were computed. (The {\it HST}
filters are of particular interest given that comparisons of isochrones with
WFC3 observations of GCs are presented in the penultimate section of this
paper as well as in Paper II.)  Examples of the histograms are shown in
Figure~\ref{fig:f4}; the three vertical bars that are plotted for
each filter indicate the differences in the predicted magnitudes betweem the
{\tt s08vt1} and {\tt s08std} BCs (top panel), and between the {\tt s21std}
and {\tt s08std} BCs (bottom panel) for [Fe/H] $= -2.5, -1.5$, and $-0.5$, in
the direction from left to right.  These results assume $\log$\,$g = 4.0$, which
is approximately the gravity of TO stars, and the $\teff$s given in the figure
caption.

%The BC transformations reported by CV14 were based on 2008 MARCS models that
%assume $v_t = 2.0$ km/s, mainly because synthetic spectra for wide ranges in
%gravity, temperature, and metallicity were produced for just this particular
%value, which is roughly the mean microturbulent velocity for lower mass stars,
%considering all evolutionary states.  However, the MARCS grids for $\log$\,$g
%\le 3.0$ and $\ge 4.0$ include a significant number of models that assume, in
%turn, $v_t = 5.0$ km/s and 1.0 km/s, given that higher values are favoured for
%bright giants while lower values appear to be more appropriate for dwarf stars
%and subgiants.  
Because the 2008 MARCS models for $v_T = 1.0$ km/s and 5.0 km/s were generated
only for limited ranges in gravity, it is not possible to use the BCs based on
these models to transform entire isochrones from the MS to the RGB tip to the
observed plane.  Nevertheless, the differences in the BCs caused by a decrease
in $v_T$ from 2.0 km/s to 1.0 km/s or an increase from 2.0 km/s to 5.0 km/s can
be evaluated for the portions of isochrones with $\log\,g \ge 4.0$ or $\le 3.0$,
respectively.  For instance, according to Fig.~\ref{fig:f4}, stellar models for
[Fe/H] $= -0.5$, [$\alpha$/Fe] $= 0.2$, $\log g$\,$g = 4.0$, and $\teff =
5576$~K will be brighter in $U$ and $B$ by $\approx 0.04$ and $\approx 0.01$
mag, respectively, and they will therefore have bluer $U-B$ colours by $\approx
0.03$ mag, if the transformations to $U$ and $B$ are derived from MARCS models
for $v_T = 1.0$ km/s instead of those for $v_T = 2.0$ km/s.  The magnitude and
colour differences are clearly much smaller at lower metallicities or in the
case of redder filters, where the magnitude differences are in the opposite
sense.  Not unexpectedly, the dependencies of $F336W$ and $F438W$ magnitudes on
$v_T$ are quite similar to those predicted for $U$ and $B$.  The same can be
said of the results for other filters that are located at similar wavelengths,
such as $F606W$ and $V$ or $F814W$ and $I_C$.  Although not shown, an increase
in $v_T$ from 2.0 km/s to 5.0 km/s gives rise to {\it fainter} $U$, $B$,
$F336W$, and $F438W$ magnitudes by up to a few tenths of a magnitude, depending
on the assumed stellar properties; see e.g., CV14, their Figs.~3 and
4\footnote{Note that the $\Delta$(mag) values plotted in the two figures given
by CV14 are in the sense mag$(v_T = 2)$ $-$ mag$(v_T = 1)$ (blue points)
and mag$(v_T = 2)$ $-$ mag$(v_T = 5)$ (red points), which is opposite to the
convention adopted in our Fig.~\ref{fig:f4}.} and the associated discussion.)

\begin{figure*}
\includegraphics[width=\textwidth]{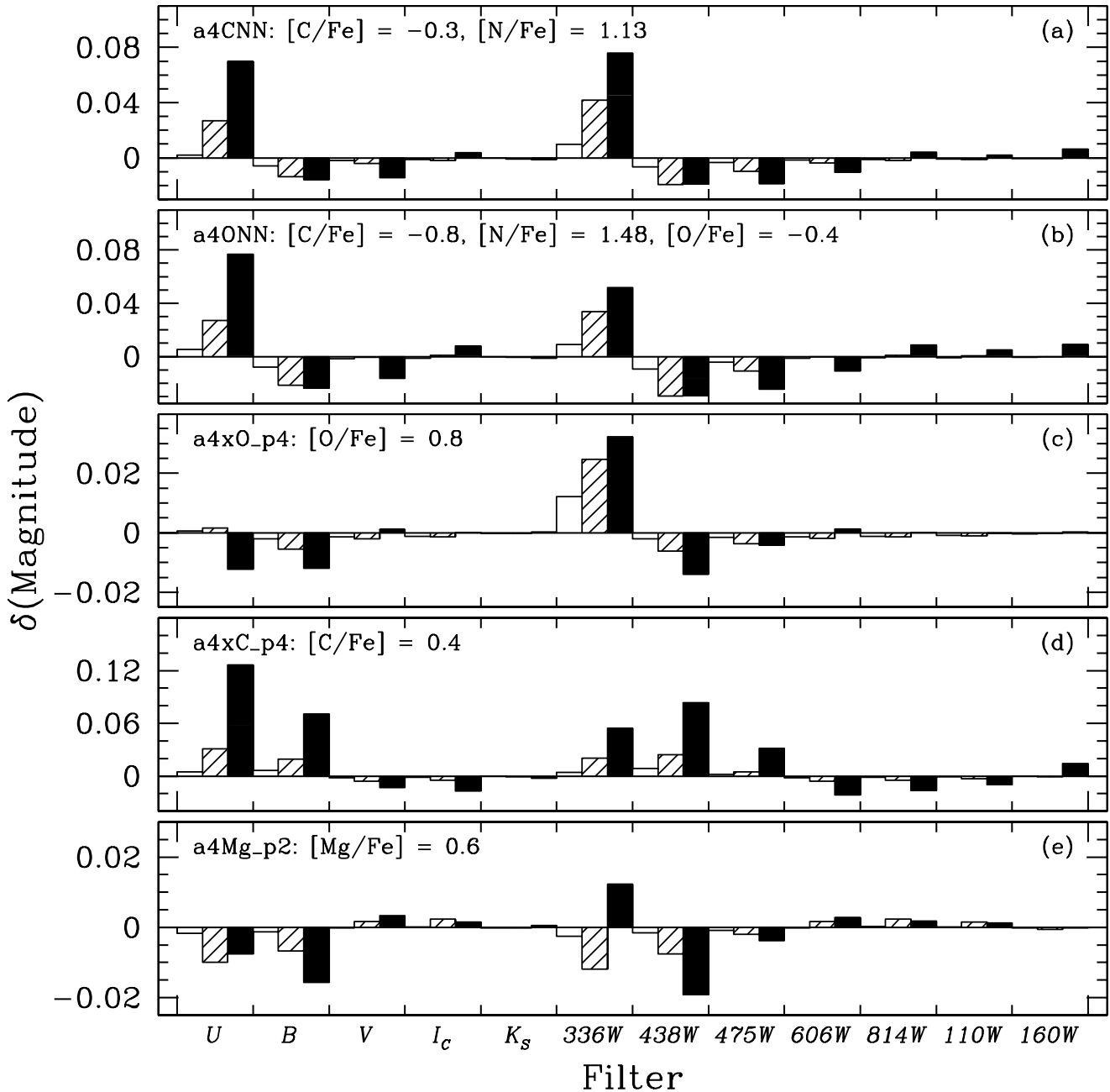}
\caption{Similar to the previous figure, except that the BCs for the reference
{\tt a4s21} mix have been subtracted from those for the indicated mixtures.
The models assume $\log$\,$g = 3.0$ and $\teff = 5283, 5166,$ and 4831~K, as
predicted by 12.5 Gyr isochrones for [Fe/H] $= -2.5, -1.5,$ and $-0.5$,
respectively, at the selected gravity.  Note that there are differences in the
ordinate scales of the various panels.}
\label{fig:f5}
\end{figure*}

\begin{figure*}
\includegraphics[width=\textwidth]{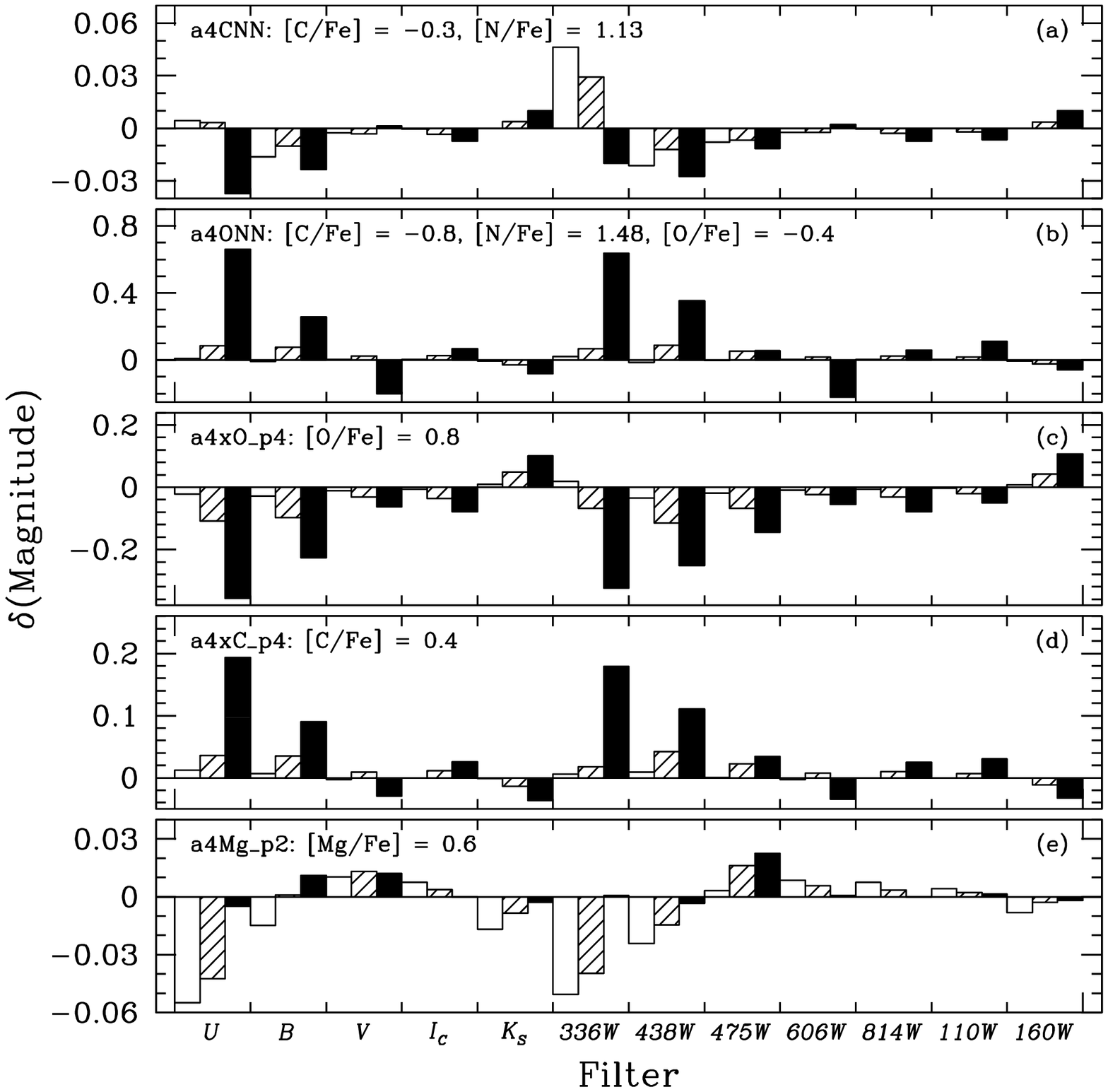}
\caption{Similar to the previous figure, except that the models assume
$\log\,g = 5.0$ and $\teff = 4170, 3828,$ and 3365~K, as predicted by 12.5
Gyr isochrones for [Fe/H] $= -2.5, -1.5,$ and $-0.5$, respectively, at the
selected gravity.}
\label{fig:f6}
\end{figure*}

Interestingly, the bottom panel of Fig.~\ref{fig:f4} shows that the recent
improvements to MARCS synthetic spectra imply brighter magnitudes at blue and
UV wavelengths than those derived from the \citet{gee08} spectra.  Note that
the actual magnitude differences due solely to the improved treatment of the
blanketing are less than those shown because the {\tt s21std} and {\tt s08std}
BCs assume different microturbulent velocities.  This can be corrected by
subtracting the predicted magnitude differences in the upper panel from those
plotted in the lower panel, resulting in, e.g., $\delta(U) \approx 0.08$ mag and
$\delta(F336W) \approx 0.09$ mag at [Fe/H] $= -0.5$.  In the case of $V$,
$F606W$, and redder filters in both photometric systems, the BCs and synthetic
magnitudes based on the current MARCS models are nearly identical with those
determined from the Gustafsson et al.~computations.  As shown in
\S~\ref{sec:6496} and Paper II,
%isochrones provide much better fits to UV photometry of GCs if they employ the
%updated BCs instead of the CV14 transformations, which are based on the older
observations appear to provide unambiguous support for
the application of systematic corrections to the CV14 BCs such as those implied
by the lower panel of Fig.~\ref{fig:f4} --- including the reduction in $v_T$ for
dwarf stars.

Histogram plots also provide a good way of visualizing the effects on a
spectrum of varying the mixture of the metals.  Figure~\ref{fig:f5} is similar
to the previous figure except that the latest BCs for the standard
[$\alpha$/Fe] $= 0.4$ mix (i.e., the {\tt a4s21} case) have been subtracted
from the BCs that are obtained on the assumption of a number of chemical
abundance variations (specifically, {\tt a4CNN}, {\tt a4ONN}, etc.), as
identified in each panel.  This particular plot assumes $\log$\,$g = 3.0$, which
is characteristic of stars that are located on the lower RGB, and the
temperatures derived from 12.5 Gyr isochrones for [Fe/H] $ = -2.5, -1.5$, and
$-0.5$ at this gravity.  According to panel (a), enhanced N (and CN) abundances
mainly affect the flux at short wavelengths (due to increased strengths of CN
bands; see \citealt{ssw11}.).  For instance, lower RGB stars with [Fe/H] $=
-0.5$ and [$\alpha$/Fe] $=0.4$ are expected to have increased $U$ magnitudes by
$\delta(U) = - \delta($BC$_U) \approx 0.07$ mag, as compared with $\delta(U)
\approx 0.03$ mag for stars with Fe/H] $= -1.5$ at a similar evolutionary state.
Since the effects on $B$ and $F438W$ magnitudes are in the opposite sense,
enhanced CN is predicted to cause significantly redder $U-B$ and
$m_{F336W}-m_{F438W}$ colours.  Such enhancements apparently have relatively
minor consequences for the IR magnitudes of stars that are just beginning their
ascent of the RGB.

If most of the C and O are converted to N as the result of
ON-cycling, the resultant BCs are qualitatively and quantitatively quite
similar to those found for the {\tt a4CNN} case; compare the results in panel 
(b) with those shown in panel (a).  Because the $U$ and $F336W$ passbands
include an NH band, the respective magnitudes are especially dependent on the
N abundance.  However, they also have some sensitivity to the abundance of
oxygen (see panel c), and differences in this dependence are presumably
responsible for the variations between panels (a) and (b) of the $\delta$(mag)
values for these two filters.  (MARCS synthetic spectra show that OH and NH are
prominent in the $F336W$ passband, along with CN in the reddest part of it.)
Note that, even if oxygen is enhanced by as much as 0.4 dex (panel c), the
consequences of such a large enhancement on the UV, optical, and IR magnitudes
and colours of lower RGB stars are relatively minor.

A 0.4 dex increase in the abundance of carbon, as illustrated in panel (d), is
much more consequential.  At [Fe/H] $> -1.5$, high C results in fainter $U$ 
and $B$ magnitudes by a few to several hundredths of a magnitude (up to $\sim
0.12$ mag at [Fe/H] $\sim -0.5$) along with appreciably redder $U-B$ and $B-V$
colours.  Interestingly, carbon affects the magnitudes measured by the $F438W$
passband more than those derived from shorter or longer wavelength filters, such
that, e.g., $m_{F336W}-m_{F438W}$ colours become bluer, while
$m_{F438W}-m_{F606W}$ colours become redder, as the result of increased C
abundances.  Fig.~\ref{fig:f5}d also shows that Mg, which is an important
contributor to the opacity in the atmospheric layers of lower RGB stars (see
the plots provided by \citealt[their Figs.~7--11]{vbd12}), is much less
important for the BCs of such stars at fixed values of $\log\,g$ and $\teff$
than C or N.  This panel indicates that a 0.2 dex enhancement will affect
predicted Johnson-Cousins-2MASS and WFC3 magnitudes at the level of $\lta
0.01$--0.02 mag.

However, these results cannot be taken at face value because giants with 
enhanced abundances of Mg (and/or Si) are predicted to be cooler (see, e.g.,
\citealt{vbd12}), which must be taken into account when stellar models are
transposed to observational CMDs.  Variations in the abundances of C, N, and O
are not expected to affect the temperatures of lower giant-branch stars because
their electrons are quite tightly bound; i.e., they have high ionization
energies.  Hence, it is necessary to take into account the impact of the
assumed chemical abundance variations on the $\teff$s of stars when evaluating
how such variations alter their CMD locations.  Before investigating this
aspect of the problem (in the next section), it is worthwhile to consider the
gravity dependence of such results as those shown in Fig.~\ref{fig:f5}.

In the case of (warmer) turnoff stars, which have gravities close to $\log$\,$g
= 4.0$, the corresponding histogram plot, which is not included here because it
is relatively simple and easy to describe in a few sentences, closely resembles
Fig.~\ref{fig:f5}, except that the derived $\delta$(mag) values are much
smaller.  At [Fe/H] $\le -1.5$, they amount to only a few thousandths of a
magnitude; consequently, the magnitudes and colours of the most metal-poor TO
stars are not significantly affected by variations in the C:N:O abundance ratio
or enhancements in the abundances of C, O, Mg, or Si. At a metallicity as high
as [Fe/H] $= -0.5$, it is only the synthetic UV magnitudes that are impacted by
$\gta 0.02$ mag.  For instance, $\delta(U) \approx +0.04, +0.04, -0.0005,
+0.03$, and $-0.003$\ mag for, in turn, the {\tt a4CNN}, {\tt a4ONN},
{\tt a4xO\_p4}, {\tt a4xC\_p4}, and {\tt a4Mg\_p2} cases, with similar findings
for $\delta(F336W)$.

More interesting is the histogram plot for $\log$\,$g = 5.0$, which is shown
in Figure~\ref{fig:f6}.  Mixtures of C and N arising from CN-cycling are not
expected to affect the broad-band magnitudes and colours of lower main sequence
(LMS) stars by more than $\sim 0.03$ mag (see panel a), though it is curious
that $F336W$ magnitudes at the lowest [Fe/H] values are fainter by 0.03--0.05
mag when the {\tt a4CNN} models generally predict somewhat brighter magnitudes
in nearly all other passbands.  As shown in panels (b) and (c), oxygen has a
dominating influence on the photometric properties of LMS stars, especially at
higher metallicities.  However, even at [Fe/H] $\sim -1.5$, reduced or enhanced
O abundances can affect predicted magnitudes at a given $\teff$\ by $\sim 0.1$
mag; note the very wide range in the ordinate values of panel (b).
According to Figs.~\ref{fig:f6}b and~\ref{fig:f6}c, the $\delta$(mag) values
for other than the $V$ and $K_S$ filters (or $F606W$ and $F160W$ in the case of
WFC3 photometry) are larger when the O abundance is reduced, and vice versa,
or when C is enhanced (panel d).  Finally, enhancements in Mg apparently have
bigger consequences for predicted magnitudes (at UV and blue wavelengths, in
particular) when [Fe/H] $= -2.5$ and $-1.5$ than at [Fe/H] $= -0.5$; see panel
(e).  However, as noted above, the translation of such findings to observed CMDs
depends on how the model $\teff$ scale is affected by the assumed metal
abundance variations.  An investigation of this issue is the next step in our
analysis.
%The same can be said of Si, though enhancements of this element are predicted
%to increase the $B$ and $F438W$ magnitudes somewhat, while reducing the $U$,
%$V$, $F336W$, and $F606W$ magnitudes (see panel e).  As a result, stars with
%increased Si abundances are expected to have bluer $U-B$, but redder $B-V$,
%colours than otherwise identical stars without such enhancements.  

\section{Isochrones for Different Mixtures of the Metals}
\label{sec:iso}

The version of the Victoria stellar evolution code that has been documented in
considerable detail by \citet[and references therein]{vbd12} has been used to
compute all of the evolutionary tracks that were needed for this project.  The
basic physics and the careful treatment of the gravitational settling of helium
incorporated in it can be regarded as state-of-the-art.  As shown by
\citet{vdc16}, the tracks produced by the Victoria code are essentially
indistinguishable from those derived from the MESA code (\citealt{pbd11}) when
as close as possible to the same physics is adopted.  The diffusion of the
metals is not considered, but this is largely inconsequential for the predicted
$\teff$\ scale, when extra mixing below surface convection zones is treated,
and for turnoff luminosity versus age relations (see the discussion by
\citealt{vbf14}).  In fact, isochrones that are generated from Victoria tracks,
using software described by \citet[and references therein]{vbd12}, match the
observed MS and RGB slopes in GC CMDs, as well as their morphologies in the
vicinity of the TO, exceedingly well (see, e.g., VBLC13). Several
studies have also demonstrated that Victoria computations do a good job
of satisfying the constraints provided by solar neighborhood stars with
well-determined distances (\citealt{vcs10}), the RR Lyrae standard 
candle {e.g., \citealt{vdc16}, \citealt{dvk17}), and binary stars in open
clusters (\citealt{bvb12}) and GCs (\citealt{bvb17}, \citealt{vd18}).     

\begin{figure*}
\includegraphics[width=\textwidth]{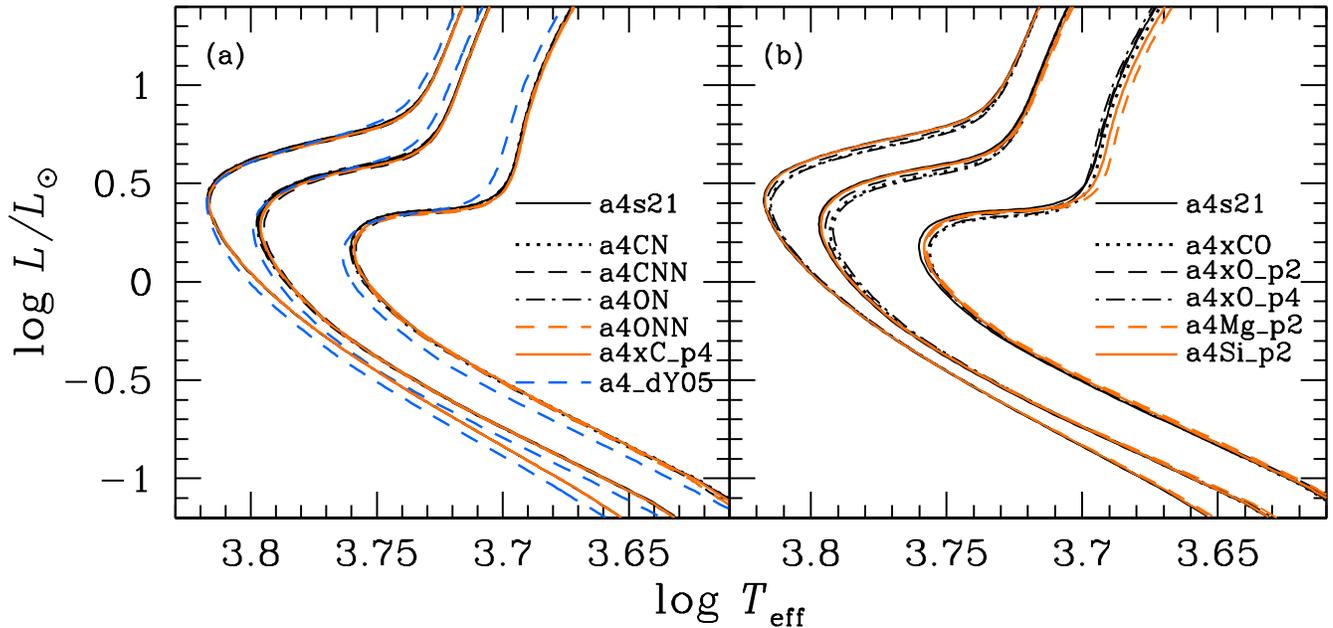}
\caption{Comparisons of 12.5 Gyr isochrones for the indicated mixtures of the
metals (see Table~\ref{tab:t1}) and [Fe/H] $= -2.5$, $-1.5$, and $-0.5$ (the
three groups of loci in the direction from left to right, respectively.
Isochrones for the standard [$\alpha$/Fe] $= 0.4$ mix ({\tt a4s21}), but with
higher $Y$ by 0.05 were taken from the study by \citet{vbf14}, and given the
name {\tt a4\_dY05}.} 
\label{fig:f7}
\end{figure*}

\begin{figure*}
\includegraphics[width=\textwidth]{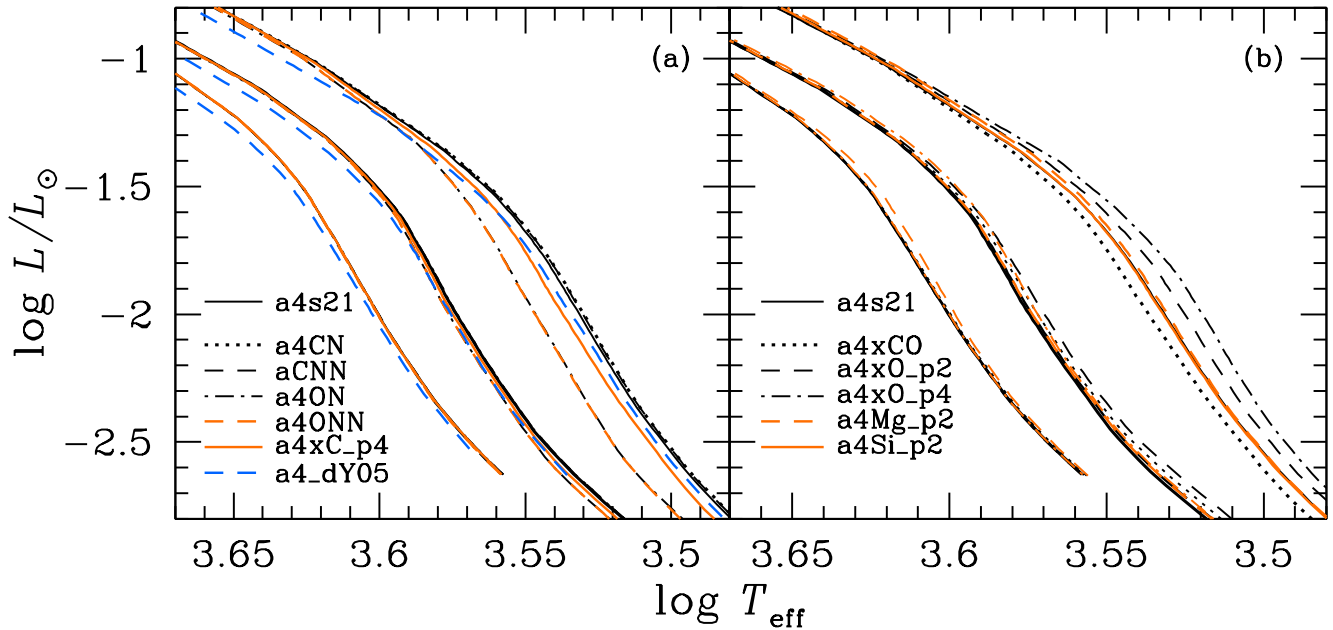}
\caption{Comparisons of the lower MS portions of the same isochrones that were
plotted in the previous figure.}
\label{fig:f8}
\end{figure*}

Moderately large grids of stellar models have already been published by
\citet{vbf14} for [$\alpha$/Fe] $= -0.4, 0.0$, $+0.4$, and the standard
variation of [$\alpha$/Fe] with [Fe/H], assuming in each case wide ranges in
[Fe/H] and $Y$.  The isochrones derived from them, which are generally referred
to as the ``Victoria-Regina isochrones", adopted the solar abundances given by 
\citet{ags09} as the reference mixture of heavy elements.  Indeed, as discussed
in \S~\ref{subsec:mix}, we have opted to make the same assumption here
%; i.e., the assumed abundance variations listed in 
%Table~\ref{tab:t1} have been applied to the Asplund et al.~mixture 
in order that the corresponding evolutionary
computations are part of the same family as the 2014 computations.

As in the previous investigation, opacities for each of the assumed mixtures of
the metals were obtained from the OPAL website\footnote{http://opalopacity.llnl.gov}
(see \citealt[and references therein]{ir96}) for temperatures $\gta 10^4\,$K, while they were generated
using the codes described by \citet{faa05} for lower temperatures.  Since we
are interested in the predicted temperatures, as well as the colours, of models
for different mixtures of the metals, it is worth mentioning that the properties
at $\tau = 100$ of MARCS model atmospheres for the different chemical abundance
choices were used to define the boundary conditions of our stellar interior
models at relatively high gravities (masses $\lta 0.4 {\cal M}_\odot$).  At
higher masses, it is preferable, for reasons given by \citet{vbf14}, to derive
the boundary pressure by integrating the hydrostatic equation from small
optical depths to the photosphere in conjunction with a $T$--$\tau$\ relation
based on an empirical solar atmosphere or accurate 3D models for the
atmosphere of the Sun; see \citet{pac13}.  Our stellar models employed the
$T$--$\tau$ relation derived by \citet{vp89} to represent the \citet{hm74} solar
atmosphere.  Those wanting more details concerning the Victoria code and/or the
methods used to generate evolutionary tracks and isochrones are encourged to
refer to the VandenBerg et al.~(2012, 2014) studies.

\subsection{Comparisons of Isochrones on the H--R Diagram}
\label{subsec:hr}

Figure~\ref{fig:f7} illustrates the upper MS, TO, and lower red-giant branches
of 12.5 Gyr isochrones for [Fe/H] $ = -2.5$, $-1.5$, and $-0.5$ and the
indicated mixtures of the metals.  At the lowest metallicity, the isochrones
are nearly coincident except in the vicinity of the TO and subgiant branch (SGB)
for those cases having higher [CNO/Fe] than the {\tt a4s21} mix; higher
C$+$N$+$O results in a fainter TO and SGB at a given age, and vice versa (see,
e.g., \citealt{vbd12}).  In general, the temperatures of MS or RGB stars are
not affected by variations in either the ratio C:N:O or the total C$+$N$+$O
abundance.  Furthermore, it is only at higher [Fe/H] values where the effects of
increased Mg and Si abundances become evident (see the loci labelled
{\tt a4Mg\_p2} and {\tt a4Si\_p2}) --- not just near the TO, but also along the
RGB.  As shown by VandenBerg et al., Mg and Si have quite a strong influence on
the predicted temperatures of giants because they are two of most abundant
metals that are also important electron donors (unlike, e.g., O and Ne).

It should be appreciated that Mg abundance variations associated with the 
observed Mg--Al anticorrelations in some GCs is a separate issue.  If Mg$+$Al
$= constant$, which seems to be typical of the majority of GCs (see
\citealt{cbg09b}), Mg--Al anticorrelations have no significant ramifications
for isochrones (\citealt{swf06}, \citealt{pcs09}).  In fact, it has been
discovered that [Si/Fe] is correlated with [Al/Fe] when the abundance of Mg is
reduced by a large amount, which indicates that there is leakage into $^{28}$Si
from the Mg-Al cycle (\citealt{ygn05}, \citealt{cbg12}, \citealt{cbl18}).  In
the case of NGC\,4833, for instance, \citet{cbg14} have found that a 0.5 dex
reduction in the Mg abundance is accompanied by increased abundances of Al and 
Si by about 1 dex and 0.2 dex, respectively, which is close to the expected
abundances if Mg$+$Al$+$Si $= constant$. Indeed, if this condition is satisfied,
isochrones that allow for different abundances of Mg, Al, and Si will be nearly
identical with those computed for the standard mix (see \citealt{cmp13}) given
that the effects of higher Al and Si abundances on low-temperature opacities
will mostly compensate for those arising from the reduced abundances of Mg.
(There should be some differences in the opacity when Mg is converted to Al and
Si because Al, in particular, has relatively few atomic lines, but they are
likely too small to have significant effects on predicted $\teff$s at low
[Fe/H] values.)

Perhaps the most striking result in Fig.~\ref{fig:f7} is that a higher He
abundance by $\delta\,Y = 0.05$ has larger consequences for the temperatures
of MS and RGB stars at a fixed luminosity than nearly all of the metal
abundance variations that we have considered.  Only 0.2 dex enhancements of
Mg and Si at only the highest metallicities, [Fe/H] $\gta -0.5$, have comparable
effects on the predicted temperatures along the giant branch.  The models for
higher $Y$, which have been given the name {\tt a4\_dY05}, were taken from the
study by \citet{vbf14}, as they were generated for the same relative abundances
of the metals as in the {\tt a4s21} mix, using exactly the same version of the
Victoria stellar evolution code that has been employed in this study.  (They
are not explicitly listed in Table~\ref{tab:t1} because BCs were not calculated 
specifically for this case.  As already mentioned, BCs have a very weak 
dependence on the assumed He abundance; consequently, the same BCs can be safely
adopted for both the {\tt a4s21} and {\tt a4\_dY05} chemical abundance mixtures.)
Worth noting is the prediction that, at a given age, differences in $Y$ affect
the slope of the SGB portion of an isochrone, but not its mean luminosity, which
depends primarily on [CNO/Fe].

\begin{figure*}
\includegraphics[width=\textwidth]{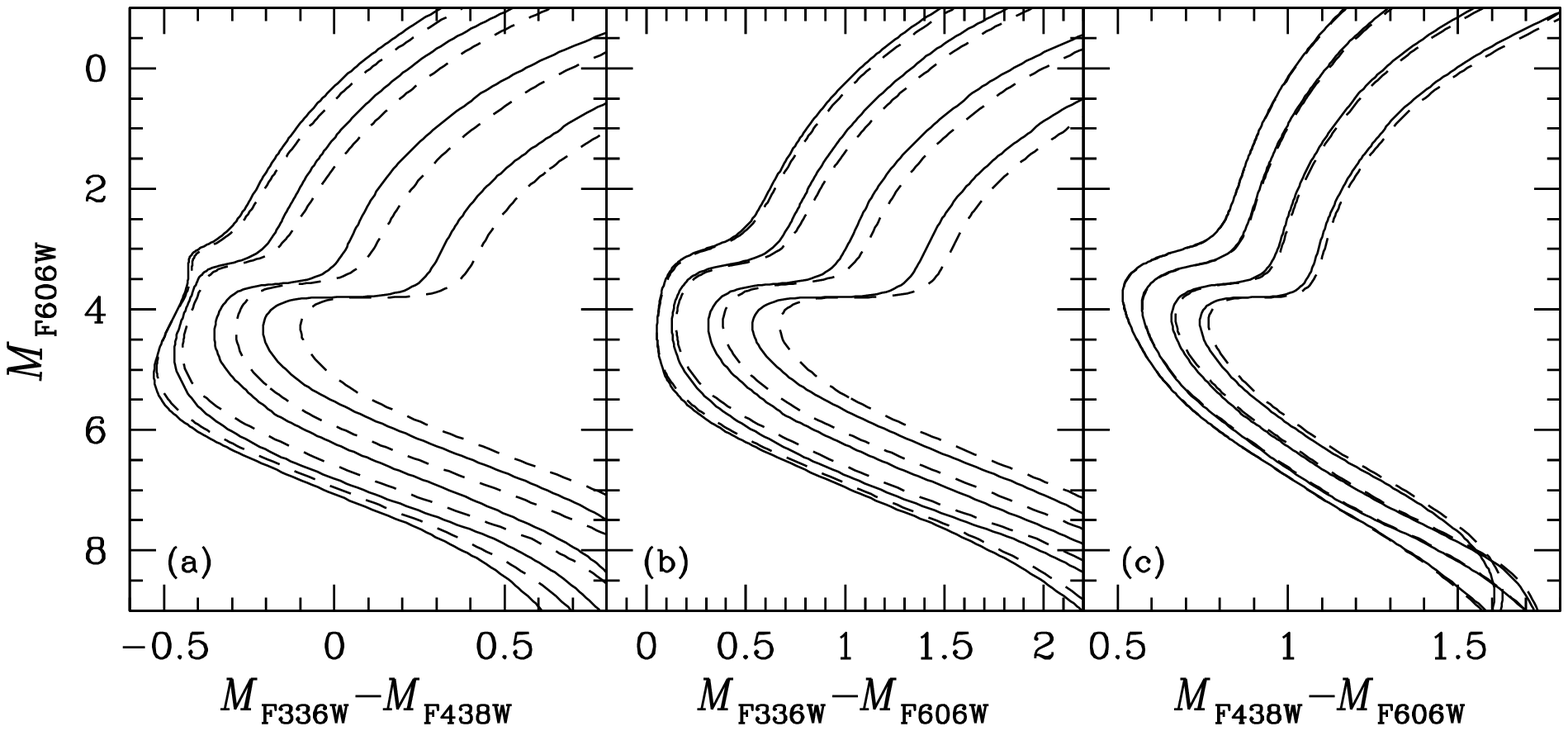}
\caption{Comparison of 12.5 Gyr isochrones for [$\alpha$/Fe] $=0.4$ and [Fe/H]
$= -2.0$, $-1.5$. $-1.0$, and $-0.5$ that have been transformed to three CMDs
involving WFC3 filters using the BCs reported by CV14 (dashed curves) and the
BCs calculated for this study (solid curves).  In the case of the
$M_{F606W}-M_{F814W}$\ colour index (not shown), the differences between the
solid and dashed loci are less than half of those shown in the right-hand
panel for $M_{F438W}-M_{F606W}$ at the same metallicities.}
\label{fig:f9}
\end{figure*}

At low masses along the MS (${\cal M}/{\cal M}_\odot \lta 0.4$), where the
blanketing of stars is strongly affected by molecules, the predicted $\teff$
scale depends much more on the mixture of heavy elements than along the upper
MS; see Figure~\ref{fig:f8}.  Furthermore, C and O clearly affect the
temperatures of LMS stellar models more than He if they are more metal rich
than [Fe/H] $\sim -1.5$.  Recall that the properties of fully consistent MARCS
model atmospheres at $\tau = 100$ were used to determine the outer boundary
conditions of the stellar interior models (notably at $\log\,L/L_\odot \lta
-1.3$); consequently, the various loci represent the best predictions that we
are able to make at the present time of both the absolute and relative locations
on the H-R diagram of LMS models for different chemical compositions.

Not surprisingly, zero-age main-sequences (ZAMSs) are particularly sensitive to
the abundance of oxygen, and somewhat less so to the abundance of carbon.  As
shown in panel (a), the large reductions in the O abundance that are assumed in
the {\tt a4ON} and {\tt a4ONN} mixtures shift the respective ZAMS loci to warmer
temperatures.  Interestingly, these sequences overlay one another almost
exactly, which suggests that the LMS is not very dependent on the
abundance of nitrogen since the only difference between these two cases is that
the {\tt a4ON} models assume a lower N abundance by 0.18 dex than those for the
{\tt a4ONN} mixture (see Table~\ref{tab:t1}).  In view of these results, one can
anticipate that significantly higher O abundances will result in cooler LMS
models, which is borne out by our computations for the {\tt a4xO\_p2} and
{\tt a4xO\_p4} mixtures; see panel (b).  Carbon appears to affect LMS models in
the opposite sense; i.e., low-mass stars with enhanced C abundances are 
predicted to be warmer ---see our results for the {\tt a4xC\_p4} and {\tt a4xCO}
mixtures.  The ZAMS for the latter case differs only slightly from the former,
even though it assumes a higher value of [C/Fe] by 0.3 dex, probably because of
the compensating effects of also assuming a higher O abundance by 0.2 dex, which
would tend to displace LMS stars to cooler temperatures.
%for the {\tt a4C1} mixture stands out from the others in showing a particularly
%steep luminosity decline with decreasing $\teff$.  This is apparently one of the
%consequences of replacing, in C-rich atmospheres, oxides by polyatomic molecules
%involving carbon that produce enormous numbers of lines.  

As shown in Figure~\ref{fig:f8}, the temperatures of low-mass ZAMS models for
[Fe/H] $\le -0.5$ are unaffected by variations in the abundance of Si or C:N
abundance ratios.   The bottom of the MS has some sensitivity to the abundance
of Mg, but the effects of variations in the abundances of C, O, and He are much
more important.  The {\tt a4CNN} case, which is similar to {\tt a4CN} but with
a higher N abundance, resulting in higher C$+$N$+$O (see Table~\ref{tab:t1}, is
predicted to be cooler than the others, but only marginally, even at the highest
[Fe/H] value.  These findings confirm the conclusions reached by \citet{pcs09},
from their own computations of stellar models, that standard $\alpha$-element
enhanced isochrones can be used to represent the CN-weak and CN-strong
sub-populations in GCs.  Although these sub-populations may be separated
photometrically through the use of suitable filters, that separation is entirely
a bolometric corrections effect, not an indicator of differences in $\teff$.  Of
course, this has long been known from spectroscopic work, such as that by
\citet{ccb98} who found that CN-weak and CN-strong stars in 47 Tuc overlay one
another on the $(B-V),V$-diagram from the upper giant branch to below the MS
turnoff with no discernible differences in their CMD locations.

Most of the results reported in this section confirm the conclusions reached in
earlier studies (several of which are referenced in \S~\ref{sec:intro}).  The
advantages of our investigation are that (i) unlike nearly all published work
on this subject to date, our models for different abundance variations have been
computed at constant [Fe/H], rather than at constant $Z$ (the mass-fraction
abundance of all elements heavier than helium), to facilitate quantitative
comparisons with observations, and (ii) our findings are based on fully
consistent, up-to-date atmosphere-interior models, which should result in
improved quantitative predictions of the dependence of the BCs on chemical
abundances.

\subsection{Comparisons of Isochrones on Selected WFC3 CMDs}
\label{subsec:cmd}

Before intercomparing the isochrones for different mixtures of the metals on
various CMDs, a brief examination of the observational consequences of the BCs
that have been generated for this project and those calculated by CV14 from the
synthetic fluxes published by \citet{gee08} is warranted.  Plotted in 
Figure~\ref{fig:f9} are 12.5 Gyr isochrones for [$\alpha$/Fe] $=0.4$ and [Fe/H]
$= -2.0$, $-1.5$, $-1.0$, and $-0.5$, as transformed to three different CMDs
using the CV14 BCs (dashed curves) and the updated BCs presented here (solid
curves).  This shows that the differences in the BCs (see the histogram plots
in \S~\ref{sec:bcs}) result in systematically bluer colours at shorter
wavelengths, by as much as $\sim 0.12$\ mag in colours that involve $M_{F336W}$,
but by $\lta 0.015$ mag in the case of optical colours.  Due to the increasing
complexity of spectra towards shorter wavelengths, improvements to the treatment
of atomic lines and molecular bands are bound to have the biggest impact on BCs
in the UV.

\begin{figure*}
\includegraphics[width=\textwidth]{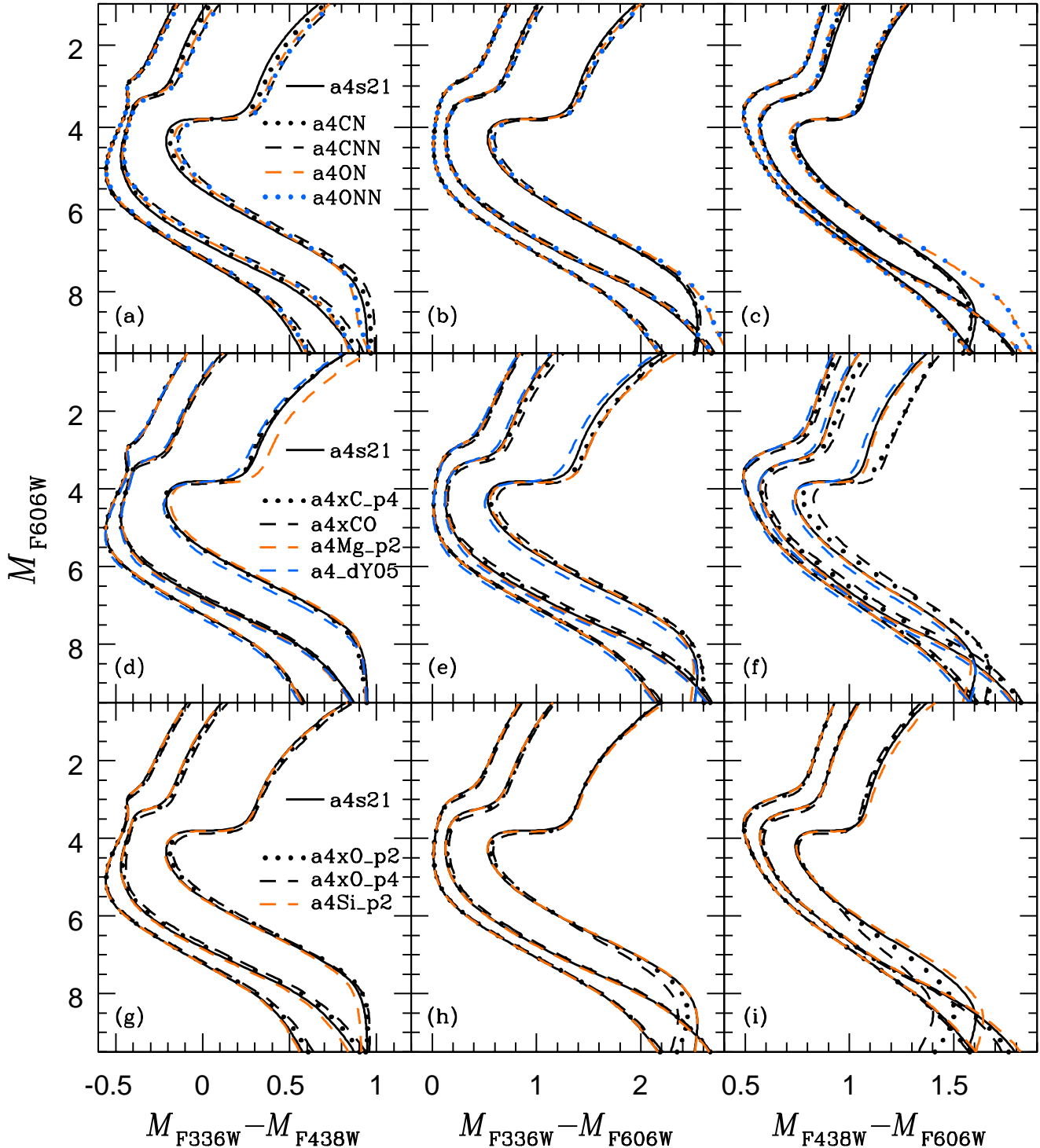}
\caption{Comparisons of 12.5 Gyr isochrones for [$\alpha$/Fe] $= +0.4$ and
[Fe/H] $= -2.5$, $-1.5$, and $-0.5$, assuming the mixtures of the metals that
are specified in the left-hand panels.  Isochrones for higher $Y$ (the 
{\tt a4\_dY05} loci) were transposed to the observed planes using the
{\tt a4s21} BCs given that bolometric corrections have very little sensitivity
to the adopted He abundance.}
\label{fig:f10}
\end{figure*}

\begin{figure*}
\includegraphics[width=\textwidth]{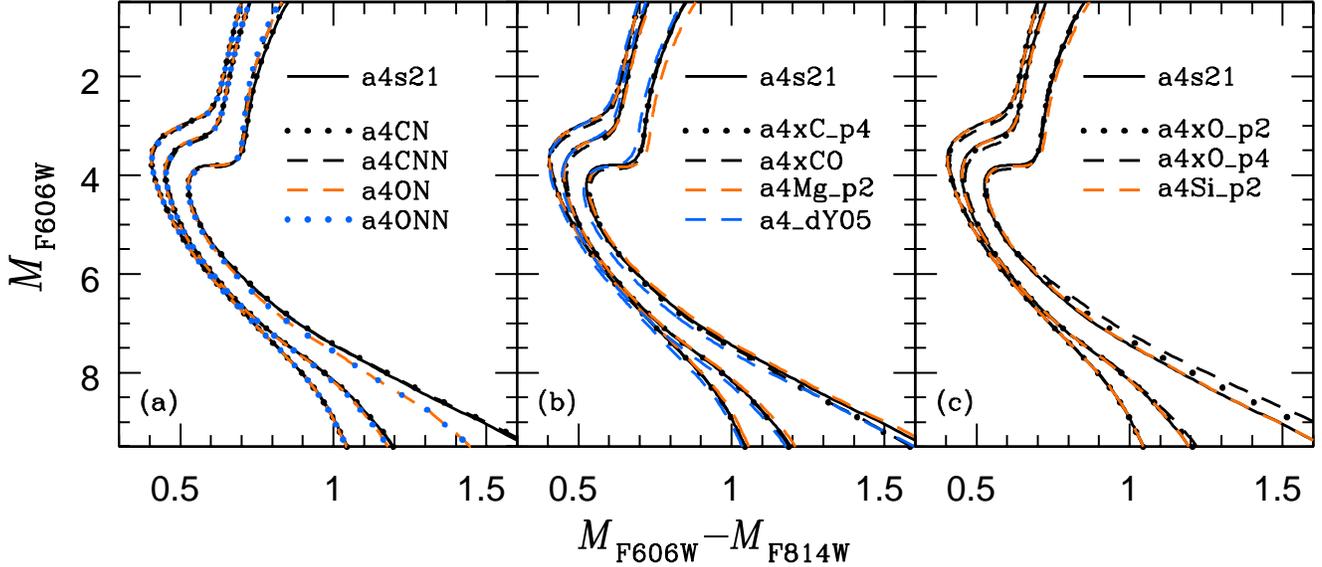}
\caption{As in the previous figure, except that the isochrones are compared on
the $M_{F606W}-M_{F814W},\,M_{F606W}$\ diagram.}
\label{fig:f11}
\end{figure*}

\begin{figure*}
\includegraphics[width=\textwidth]{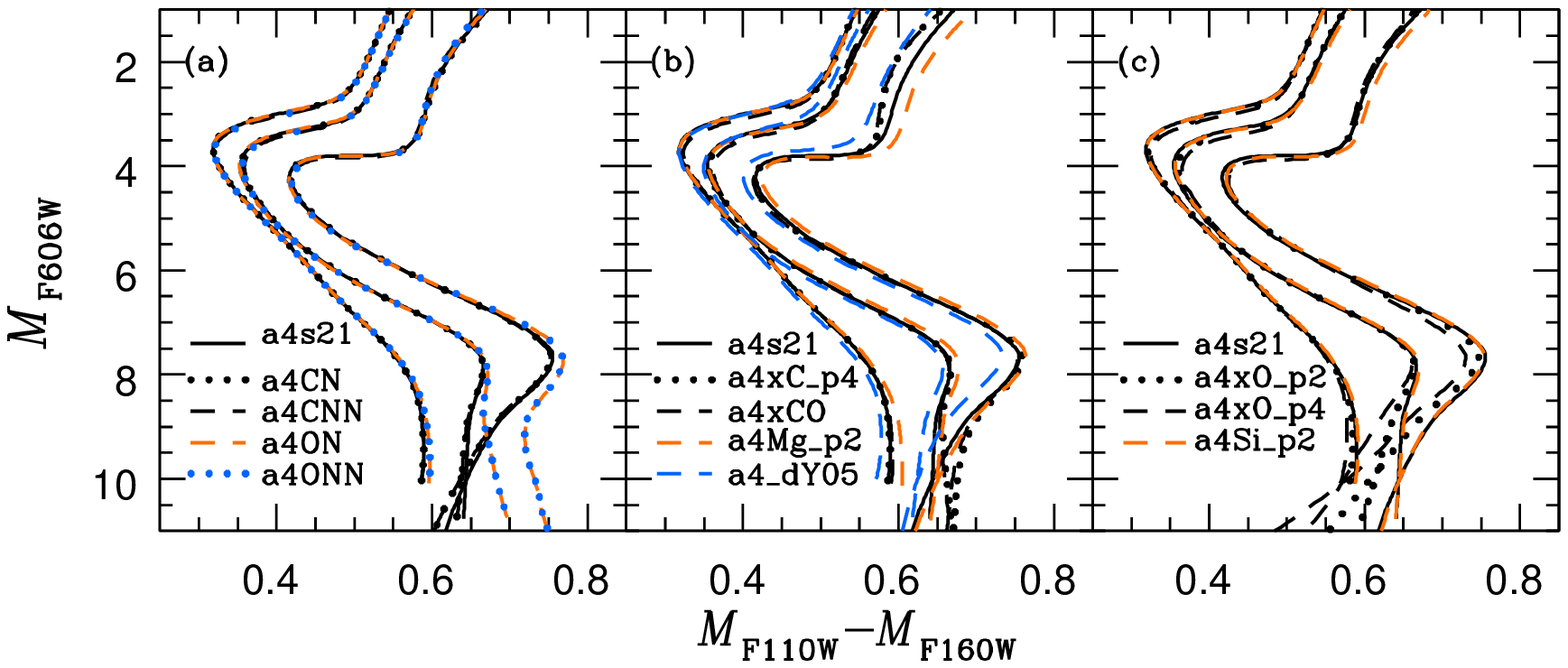}
\caption{Comparisons of the predicted IR colours along 12.5 Gyr isochrones for
[Fe/H] $= -2.5$, $-1.5$, and $-0.5$ with the indicated mixtures of the metals.}
\label{fig:f12}
\end{figure*}

The next three figures illustrate the main results of this investigation.  As
discussed by \citet{pmb15}, the $F336W$ and $F438W$ filters provide the means to
discriminate between stars with different CN strengths.  To be more specific:
because the $F336W$ passband contains an NH band, N-rich stars will have fainter
$F336W$ magnitudes than those described as CN-weak stars.  Similarly, given that
the $F438W$ passband includes CN and CH bands, C-rich stars will be fainter in
$F438W$; consequently, stars containing CNO-processed material will have redder
$m_{F336W}- m_{F438W}$ colours than those with [C/Fe] and [N/Fe] $\sim 0$.  These
expectations are confirmed by the BC histogram plots in our Fig.~\ref{fig:f5},
as well as the upper left-hand panel in Figure~\ref{fig:f10}, which shows that
that the {\tt a4CN} and {\tt a4CNN} isochrones, with reduced C and enhanced N
abundances, are significantly redder than those for scaled-solar abundances of
C and N (the solid curve representing the {\tt a4s21} mixture) on the
$M_{F336W}-M_{F438W},\,M_{F606W}$ diagram.  Note that the isochrones for the
{\tt a4ON} and {\tt a4ONN} mixtures, which assume somewhat higher N abundances,
are displaced to even redder colours along the MS and RGB, though bluer colours
are predicted for these cases along the LMS (because low-mass stars with
depleted O abundances are predicted to be considerably warmer than those with
normal abundances ([O/Fe] $\sim 0.4$); see Fig.~\ref{fig:f8}).

Variations in the C, N, and O abundances due to CN- and ON-cycling have much
smaller consequences for the colours of MS and lower RGB stars that are derived
from the $F438W$, $F614W$, and $F814W$ passbands.  Such colours, as shown by
the isochrones plotted in Fig.~\ref{fig:f10}c and the left-hand panel of
Figure~\ref{fig:f11}, tend to be somewhat bluer, which is a direct consequence
of differences in the respective BCs (see, e.g., Fig.~\ref{fig:f5})  Although
Fig.~\ref{fig:f10}b suggests that $M_{F336W}-M_{F606W}$ is less sensitive to N
abundances than $M_{F336W}-M_{F438W}$, this impression is due entirely to the 
fact that the abscissa scale of the middle panel is compressed by a factor of
two compared with that adopted for the left-hand panel.  

The large CMD-to-CMD variations in the location of the LMS models for the
{\tt a4ON} and {\tt a4ONN} mixtures relative to those for the standard mix
({\tt a4s21}) reflect the very different $\teff$ dependencies of the BCs for 
the WFC3 filters at high gravities.  For instance, at $\log\,g \gta 4.7$, the
$\delta$(BC) values plotted in Fig.~\ref{fig:f2} for the {\tt a4ON} mixture 
(also assuming [Fe/H] $= -0.5$), vary strongly with temperature, which results
in apparently very steep functions of gravity.  Interpolations in those data
yield $\delta$(BC) values of $+0.01, -0.13, +0.13$, and $-0.02$ mag for, in
turn, the $F336W$, $F438W$, $F606W$, and $F814W$ filters at $\log\,g = 4.85$.
If the corresponding $\delta$(colour) values are calculated, one obtains the
predicted offsets of the ZAMS loci at $M_{F606W} \sim 9.4$ that have been
plotted in the top row of panels in Fig.~\ref{fig:f10} and the left-hand panel
of Fig~\ref{fig:f11}.  Since the spline fits to the $\delta$(BC) values in 
Fig.~\ref{fig:f2} must necessarily pass through the predictions for $\log\,g =
4.7$ and 5.0, the interpolations for intermediate gravities should be quite
trustworthy.  In any case, comparisons with observations, such as those
presented in Paper II, must be carried out to validate the models.

As expected, enhanced C abundances have the largest effects on
$M_{F336W}-M_{F438W}$ colours, making them considerably redder than those
predicted for scaled-solar abundances (see Fig.~\ref{fig:f10}f).  In the case
of lower RGB stars, the separation of the {\tt a4xC\_p4} models from those for
the {\tt a4s21} mix, as well as the prediction that higher C causes somewhat
bluer $M_{F336W}-M_{F438W}$ colours, follows directly from the dependence of
the BCs on the C abundance (see Fig.~\ref{fig:f5}) --- since enhanced carbon
has almost no impact on the temperatures of evolved stellar models, as shown in
Fig.~\ref{fig:f7}.  The middle row of panels in Fig.~\ref{fig:f10}, along with
Fig.~\ref{fig:f11}b, also highlight the importance of Mg, especially at higher
metallicities, and He.  For both of these elements, the shifts in the colours of
the respective isochrones are mostly a reflection of temperature offsets
relative to the {\tt a4s21} isochrone rather than to variations in the BCs due
to increased abundances.  At fixed values of $\teff$, a 0.2 dex enhancement in
the abundance of Mg affects the BCs at the level of only $\sim 0.005$--0.015
mag (see Fig.~\ref{fig:f5}).
%which is comparable to the uncertainties associated with photometric zero-points.

The bottom row of plots in Figure~\ref{fig:f10} and Fig.~\ref{fig:f11}c show
how the CMD locations of isochrones for the same age and metal abundances are
affected by altering the abundance of oxygen and silicon, the former by 0.2 and
0.4 dex and the latter by 0.2 dex.  Consistent with the implications from BC
calculations (such as those in Fig.~\ref{fig:f5}), our models indicate that
upper MS, TO, and lower RGB stars are affected by O abundance enhancements in
only a relatively minor way --- except at higher metallicities (see panel i).
$F606W$ is probably the filter that is most sensitive to the presence of TiO
bands, which will be especially prominent in spectra of cool, metal-rich stars.
Moreover, a significant gravity dependence is expected because both the Ti and
O pressures, and hence the TiO pressure, are sensitive to the overall gas
pressure, and therefore $\log\,g$.  In the case of LMS stars, in which many
other molecules involving O form (notably H$_2$O), both the predicted $\teff$s
(Fig.~\ref{fig:f8}) and the BCs (Fig.~\ref{fig:f6}) are very dependent on the
abundance of oxygen, though not always in the same sense.  That is, even though
stellar models that allow for O abundance enhancements are cooler than those
without such enhancements, it is not necessarily the case that they are also
redder.  In fact, as shown in Fig.~\ref{fig:f10}i,
$M_{F438W}-M_{F606W}$ colours are bluer, whereas $M_{F606W}-M_{F814W}$ colours
(see Fig.~\ref{fig:f11}c) are redder if O has increased abundances.

\begin{figure*}
\includegraphics[width=\textwidth]{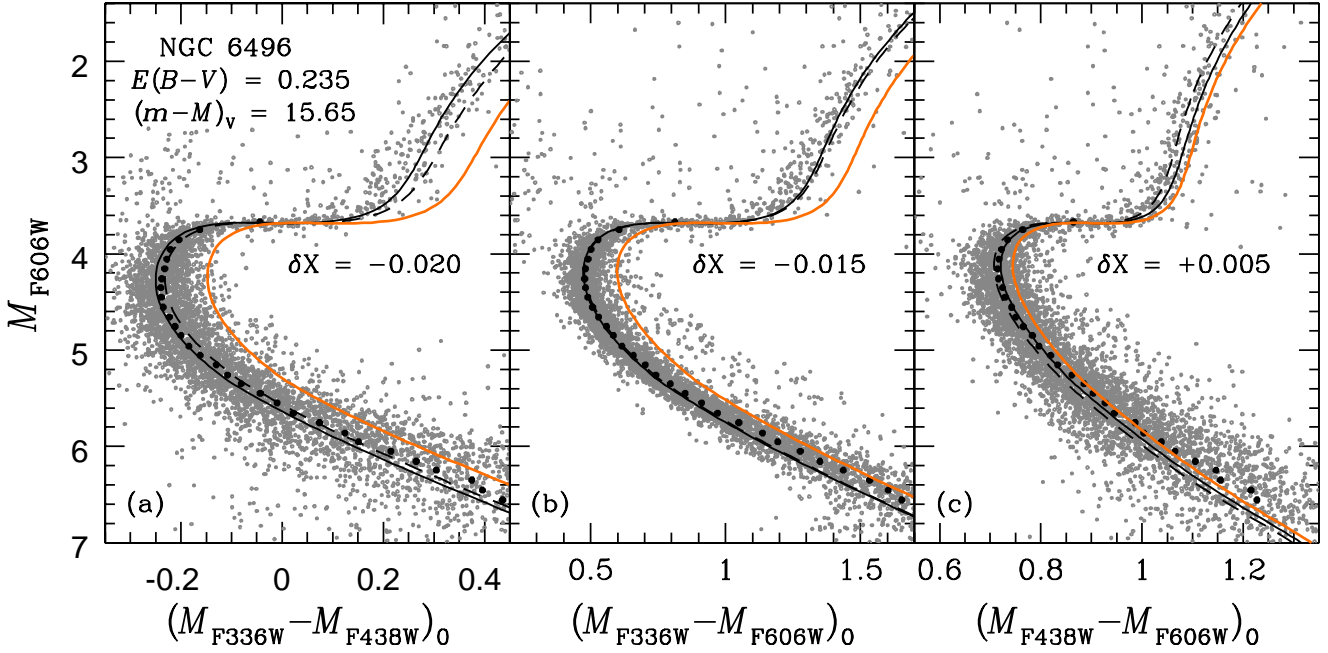}
\caption{Fits of 11.0 Gyr isochrones for [Fe/H] $= -0.5$, $Y= 0.26$, and
[$\alpha$/Fe] $= 0.4$ to WFC3 observations of NGC\,6496 (from \citealt{nlp18})
on the assumption of the indicated reddening and apparent distance modulus.
Isochrones for the {\tt a4s21} and {\tt a4CN} mixtures, using the BCs that were
computed for this project, are represented by the solid and dashed loci, in
black, respectively.  If the CV14 BCs are used instead to transform the 
{\tt a4s21} isochrone to the observed planes, one obtains the solid, orange
curves.  The isochrone colours were adjusted by the amounts given by the
$\delta\,X$ values in each panel; these adjustments are needed to obtain
satisfactory fits of the black isochrones to the cluster fiducial sequences
(filled circles) in the vicinity of the turnoff.  Note that, for any colour
index $\zeta-\eta$, the colour excess can be calculated from $E(\zeta-\eta) =
(R_\zeta-R_\eta)E(B-V)$, using the tabulated values of $R_\zeta$ and $R_\eta$
reported by CV14 (see their Table A1).  Information given in the same source
provides the means to convert $(m-M)_V$ to $(m-M)_{F606W}$.}
\label{fig:f13}
\end{figure*}

%As Paper III contains a fit of isochrones to IR data for $\omega\,$Cen, we
We conclude this section with an intercomparison and brief discussion of the
predicted $M_{F110W}-M_{F160W}$\ colours of stellar models.  Plotted in
Figure~\ref{fig:f12} are isochrones for the same ages, metallicities, and the
various chemical abundance mixtures, as indicated, that were tbe subject of the
two previous figures.  The effects on the colours of varying the abundances of
He, C, O, Mg, and Si, as well as the ratio C:N:O and [CNO/Fe], are all clearly
displayed.  MS and RGB stars with enhanced Mg and Si are cooler along the giant
branch, resulting in redder IR colors, particularly at higher metallicities,
while the reverse is true if they have higher $Y$.  As expected from the
relevant BCs in Fig.~\ref{fig:f5}, giants with increased C abundances are
predicted to have bluer $M_{F110W}-M_{F160W}$ colours.  However, oxygen clearly
plays a major role in determining the colours of LMS stars in the IR, as already
pointed out by \citet{mmb19} in their study of NGC$\,$6752.  If the LMS stars in 
GCs span a wide range in the $m_{F110W}-m_{F160}$ colour at a given magnitude,
the bluest ones are likely to have oxygen abundances that are close to initial
abundances (i.e., when they formed), while the reddest ones are likely to be
members of the cluster's most oxygen-poor population, though they could also
be C-rich stars; see Fig.~\ref{fig:f12}b.

\section{NGC\,6496: An Initial Test Case}
\label{sec:6496}

Since the effects of metal abundance variations on the colours of stars 
become larger with increasing [Fe/H], mainly because (i) there are more atoms of
the metals at higher metallicities and (ii) more metal-rich stars at a similar
evolutionary state are cooler, one might expect that WFC3 observations of the
most-metal rich GCs will present the greatest challenge for stellar models.
Indeed, to obtain a first impression of the capabilities of our improved BCs
over those given by CV14, which are based on the previous generation
of MARCS model atmospheres and synthetic spectra, we decided to fit isochrones
to observations of NGC\,6496.   According to \citet{cbg09a}, this cluster has
[Fe/H] $= -0.46$, making it one of the best available targets for the fitting of
the highest metallicity ([Fe/H] $ = -0.5$) isochrones that have been computed
for mixtures of the metals in the ``a4" series.  

WFC3 photometry reported by \citet{nlp18} was obtained via the website that
they provide.  The CMD
%\footnote{http://groups.df.unipd.it/ESPG/treasury.php} The CMD
was limited to stars with membership probabilities $\gta 98$\%, photometric
errors $< 0.02$ mag, and quality-of-fit ({\tt QFIT}) parameters $\gta 0.99$.
Because we are primarily interested in the upper MS, TO, and lower
RGB stars, we opted to use their so-called ``Method 1" photometry.
The MS and TO observations were sorted into 0.1 mag bins in
$m_{F606W}$, and median fiducial points were determined for each bin.  Fitting
isochrones to median fiducial sequences has the important advantage that
subjective errors have little or no impact on determinations of the best
estimate of the cluster age that corresponds to an adopted distance modulus.

Figures~\ref{fig:f13} and~\ref{fig:f14} show that isochrones for [Fe/H]
$= -0.5$, $[\alpha$/Fe] $= 0.4$, $Y = 0.26$, and an age of 11.0 Gyr provide a
far superior fit to the photometry of NGC\,6496, particularly the UV
observations, when they are transformed to the various CMDs using the improved
BCs (the black loci) rather than those given by CV14 (the orange curve).  These
plots assume $E(B-V) = 0.235$, which is the line-of-sight reddening according to
the \citet{sfd98} dust maps.  Indeed, $E(B-V) = 0.22$--0.24 has also been found
in several previous spectroscopic and photometric studies (e.g., \citealt{fg91},
\citealt{pdc03}, \citealt{alg15}, VBLC13).  The adopted apparent distance
modulus is supported by the fit of a zero-age horizontal branch (ZAHB) locus to
the observed HB stars (see VBLC13), and there is ample evidence that distance
moduli which are derived in this way are in excellent agreement with empirical
distance determinations based on the RR Lyrae and local subdwarf standard
candles (see e.g., \citealt{vbf14}, \citealt{vd18}).  Assuming that there has
been some enrichment of helium since the Big Bang, it is reasonable to adopt $Y
= 0.26$ for NGC\,6496.  Whether or not a somewhat lower or higher value of $Y$
is more appropriate for this GC is, anyway, a moot point because small changes
to the assumed He abundance will have only minor consequences for the fits to
the photometric data. 
 
Both of the solid curves assume the {\tt a4s21} metal abundance mixture, while the
dashed isochrone assumes the {\tt a4CN} mix.  On the assumption that NGC\,6496
has comparable numbers of CN-weak and CN-strong stars, the isochrones have been
adjusted horizontally in color by the indicated $\delta\,$X values so that the
black loci straddle the median fiducial sequences.  Approximately 20\% of the
differences along the MS between the solid curves in black and orage are due to
the reduction in the micro-turbulent velocity from 2 km~sec$^{-1}$, as assumed
in the CV14 transformations, to 1 km~sec$^{-1}$, which is the preferred value
for dwarf stars in both previous and current MARCS models.  As our main purpose
in considering NGC\,6496 was to check how well our models are able to reproduce
observed CMDs, it is very gratifying to find that the colour offsets between
theory and observations are small, certainly within the uncertainties
associated with the cluster properties (metallicity, reddening), the model
$\teff$\ scale, and the zero-points of both the photometry and the computed BCs.
The relatively large scatter in the colours of turnoff stars is presumably due
mostly to photometric errors, with some contributions due to the presence of
binaries, perhaps modest star-to-star He abundance variations, and possibly 
some amount of differential reddening.  As discussed in \S~\ref{sec:bcs},
variations in C:N:O, at constant C$+$N$+$O, are predicted to have very little
effects on TO colours.

\begin{figure}
\includegraphics[width=\columnwidth]{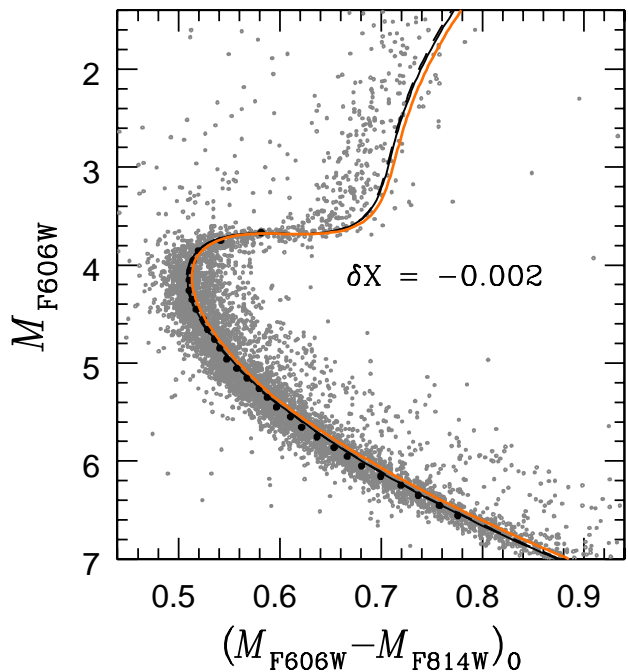}
\caption{As in the previous figure, except that the isochrones are fitted to
$F606W,\,F814W$ photometry.} 
\label{fig:f14}
\end{figure}

One unexpected result of fitting isochrones to WFC3 observations of NGC\,6496
is that the predicted and observed $M_{F606W}-M_{F814W}$\ colour of the TO are
in excellent agreement, as indicated by the small value of $\delta\,X$ in
Fig.~\ref{fig:f14}.  This is surprising because VBLC13 found that the same
Victoria-Regina isochrones, when compared with {\it HST} Advanced Camera for
Surveys (ACS) data for this and $\sim 50$ other GCs (as obtained by
\citealt{sbc07}), generally required a blueward shift of $\approx 0.02$\ mag to
reproduce observed turnoffs, independently of the cluster under consideration.
Although the $F606W$ and $F814W$ filters in the ACS and WFC3 instruments are not
identical, ACS $M_{F606W}-M_{F814W}$ colours for MS stars appear to be just
slightly bluer, by $\approx 0.003$ mag, than the corresponding WFC3 colours,
based on our examination of the CV14 BCs.  However, similar comparisons of
observed ACS and WFC3 CMDs for the same GC (from \citealt{sbc07} and NLP18,
respectively, reveal differences amounting to $\approx 0.018$ mag, but in the
opposite sense.  Clearly, there is a net zero-point difference of about 0.02 mag
between predicted and observed $M_{F606W}-M_{F814W}$\ colours in the ACS and
WFC3 systems.  Regardless of whether this is a problem with the BC transformations
or with the photometric zero-points, very little, if any, offsets to WFC3
$M_{F606W}-M_{F814W}$ colours for the MS stars in GCs appear to be necessary.

As far as we are aware, \citet{mcc17} were the first researchers to fit
isochrones to UV-optical CMDs derived from WFC3 observations of GCs
(specifically NGC$\,$6752).  Using BCs based on Kurucz ATLAS12 model atmospheres
(\citealt{ku14}) and synthetic spectra produced by the SYNTHE code
(\citealt{ku05}), they found that their models reproduced the observed
$F606W,\,F814W$ photometry of turnoff stars quite well, but that they are too
red along the giant branch --- which is qualitatively similar to our results for
NGC\,6496 (see Fig.~\ref{fig:f14}), as well as the fits to ACS $F606W,\,F814W$
photometry (\citealt{sbc07}) reported by VBLC13 for a few dozen GCs.  (The cause
of this $\sim 0.03$--0.04~mag discrepancy along the lower RGB is not known at the
present time, though we will revisit this problem in Paper II.)  Interestingly,
Martins et al.~found that the predicted $M_{F336W}-M_{F814W}$ colours are too
blue for both TO and RGB stars by quite large amounts ($\sim 0.12$--0.15 mag,
depending on the assumed value of $Y$).  Apparently Kurucz models predict too
much flux at short wavelengths, whereas our improved MARCS models, even in the
case of clusters that are more metal rich by $\sim 1$~dex, are not obviously
deficient in this regard (see Fig.~\ref{fig:f13}).

The encouraging results that we have obtained for NGC\,6496 gives us considerable
optimism that our models will fare quite well in applications to other systems
which have been much more thoroughly studied insofar as their chemical
properties are concerned.  In Paper II, we analyze the CMDs derived from
WFC3 observations of six GCs (47 Tucanae, NGC\,6362, M\,5, M\,3, M\,55, and
M\,92) that span a range in [Fe/H] from $\sim -0.7$ to $\sim -2.4$.

\section{Summary}
\label{sec:sum}

This investigation was undertaken primarily to evaluate the consequences of
metal abundance variations on the magnitudes and colours of stars to complement
the study by \citet{vbd12}, who determined the effects on predicted luminosities
and $\teff$s of increasing the abundances of several of the most abundant metals,
in turn, by 0.4 dex.  It was also our intention at the outset to apply our
stellar models to the photometry of a few GCs from the recent {\it HST} UV
Legacy Survey (\citealt{pmb15}, \citealt{nlp18}) in order to assess how well
they are able to reproduce observed colours from the UV to the IR over a wide
range in [Fe/H] and, if possible, to refine our understanding of these systems.
The project as a whole has been divided into two parts: this paper, which
focuses on the model computations, and Paper II, which compares our isochrones
with the observed CMDs of 47 Tuc, NGC\,6362, M\,5, M\,3, M\,55, and
M\,92. 
%Paper III applies our stellar models to the discrete multiple stellar
%populations in $\omega$\,Cen that have been identified by \citet{bma17}.

This work began with the definition of a dozen mixtures of the metals that 
assumed either enhanced abundances of individual elements (C, O, Mg, or Si),
or variations in the C:N:O abundance ratio for different values of [CNO/Fe].
% or
%extreme values for the O--Na or Mg--Al anticorrelations.
For each of the
adopted mixtures, we computed (i) MARCS model atmospheres and high-resolution
synthetic spectra for suitable ranges in $\log$\,$g$, $\teff$, and [Fe/H],
(ii) BCs from these spectra for many of the broad-band filters currently in
use, including, but not limited to, those for the 2MASS, SDSS, Johnson-Cousins,
and the {\it HST} ACS and WFC3 photometric systems, and (iii) grids of
evolutionary tracks for, in most cases, masses in the range $0.10 \le
{\cal M}/\msol \le 1.0$, from which $\sim\,7$--14 Gyr isochrones could be
generated.  For masses $\lta 0.4 \msol$, the boundary conditions of the stellar
interior models were derived from the MARCS model atmospheres at an optical
depth $\tau = 100$ in order to make the best possible predictions of stellar
temperatures along the lower MS (see the discussion of this point by
\citealt{vbf14}).  A scaled solar $T$--$\tau$ was used to describe the
atmospheric structure at higher masses.  Importantly, all of the
model atmospheres, the synthetic spectra, and the stellar models were computed
for exactly the same mixtures of the heavy elements.

Worth emphasizing is that the current version of the MARCS spectral synthesis
code contains a number of improvements over the version that was used for the
grids published by \citet{gee08}.  In particular, it incorporates an improved
treatment of C, N, and O (notably of molecules formed out of these atoms), with
important consequences for UV spectra, irrespective of variations in the 
abundances of these elements are considered.  That is, the updated BCs for the
[$\alpha$/Fe] $ = 0.4$ reference mixture differ from those derived by CV14 from
2008 MARCS spectra by amounts that increase systematically with decreasing
wavelength (see Fig.~\ref{fig:f4}).  (In order to facilitate comparisons with
observations, we have produced tables of the updated BCs and the necessary
computer codes to transform Victoria-Regina isochrones for [$\alpha$/Fe] $=
0.4$, with and without the metal abundance variations that have been considered
in this investigation; they may be obtained from the web site
that is provided at the end of this paper.)

Our results have demonstrated the necessity of treating the chemical abundances
in the atmosphere, interior, and spectrum synthesis models as consistently as
possible.  In particular, the temperatures of LMS stars are predicted to
have a strong dependence on the abundance of oxygen in the atmospheric layers,
much more so than in the case of any other metal.  This dependence is in the
sense that O-rich stellar models are significantly cooler, at a given
luminosity, than those without such enhancements or those with depeleted oxygen
(see Fig.~\ref{fig:f8}).  On the other hand, higher O abundances result in,
e.g., bluer $M_{F110W}-M_{F160W}$ colours at a fixed $\teff$ and gravity
(Fig.~\ref{fig:f6}).  Indeed, enhanced O has the net effect of making the LMS
portion of an O-rich isochrone considerably bluer on the
$M_{F110W}-M_{f160W},\,M_{F606W}$ diagram than that of an otherwise identical
isochrone without the oxygen enhancement (see Fig.~\ref{fig:f12}).  Thus, 
insofar as this particular colour is concerned, O-rich stars are expected to be
bluer, despite being cooler. (As shown in Fig.~\ref{fig:f6}, enhanced
O can also result in redder colours, depending on the colour index that is
considered.)

By contrast, the abundances of Mg or Si are of relatively little consequence
for the temperatures of LMS stars (see Fig.~\ref{fig:f8}.  These elements are
also unlike oxygen in that they have a significant influence on the temperatures
of upper MS, TO, and RGB stars (see \citealt{vbd12} and our
Fig.~\ref{fig:f7}).\footnote{It is well known that the luminosities and $\teff$s
of TO stars are dependent on the abundance of oxygen {\it in their deep
interiors} because of structural changes that occur when the C$+$N$+$O
abundance is changed, but that is a separate issue.}  Enhanced Mg or Si results
in redder colours along the giant branch (see Fig.~\ref{fig:f5}), which is due
partly to their effects on the stellar $\teff$ scale and partly to their impact
on bolometric corrections.  However, modest variations in the He abundance
($\delta\,Y \lta 0.05$) have larger consequences for the temperatures of upper
MS, TO, and RGB stars, particularly at lower metallicities, than Mg and Si. 

In fact, He abundance variations, which are known to exist in most GCs, may have
more profound implications for our understanding of these systems than the
observed CN, ON, ONa, and MgAl anticorrelations.  For instance, \citet{gls13}
has found that the colours of HB stars are correlated with the abundances of
$p$-capture elements in M\,5, with decreasing O and increasing Na abundances
in the direction from red to blue.  Similar findings in M\,22 (\citealt{gls14})
and in NGC\,6723 (\citet{gls15} led these researchers to conclude that
chemical composition (primarly He) is the main driver behind the distributions
of stars along the lengths of cluster HBs.  This is not a new idea (see, e.g.,
\citealt{dc04}), and indeed modern simulations of HB populations (e.g.,
\citealt{dvk17}) provide the best matches to the observed CMD distributions of
the core He-burning stars if $Y$ varies by small ($\sim 0.01$--0.02) to moderate
amounts ($\sim 0.04$--0.05), depending on the length of the HB (especially the
length of the blue tail).  However, as already noted in \S~\ref{sec:intro}, the He
abundance variations that have been derived from chromosome maps
(\citealt{mmr18}) are sometimes in rather poor agreement with other
determinations.

We concur with the results of \citet{pcs09}, who showed that isochrones for 
upper MS, TO, and RGB stars are independent of differences in the C:N:O
abundance ratio, provided that C$+$N$+$O $= constant$.  Indeed, even along the
LMS, variations in C:N remain inconsequential, which follows from the fact that
model atmospheres predict nearly identical properties (e.g., pressure,
temperature) at $\tau = 100$ for the base {\tt a4s21} mix, which is relevant
to CN-weak stars, and for the {\tt a4CN} mixture, which has C and N abundances
that are more characteristic of CN-strong stars.  Since the MARCS atmospheres
are used as boundary conditions, the isochrones for these two cases superimpose
one another almost exactly on the theoretical plane.  As a result, there is no
need to compute evolutionary tracks and isochrones for different C:N ratios.
One can simply apply the BCs for, e.g., the {\tt a4CN} mixture to the isochrones
for the {\tt a4s21} mix for the same values of [$\alpha$/Fe] and [CNO/Fe] in
order to determine how predicted magnitudes and colours are affected by the
variations in CN.  The same point has been made by Pietrinferni et al.
%
%Interestingly, the atmospheric boundary conditions for the {\tt a4CNN} mix are
%also very similar to those for the {\tt a4s21} and {\tt a4CN} mixtures, despite
%large differences in the assumed N abundances (see Table~\ref{tab:t1}), which
%explains why the LMS portions of all three isochrones are virtually
%indistinguishable.  The apparent insensitivity of the HR-diagram locations of
%very low mass stars to the abundance of N also explains why the ZAMS loci for
%the {\tt a4ON} and {\tt a4ONN} cases are also nearly coincident, since they
%assume identical chemical compositions except for a difference of $\approx 0.2$
%dex in the abundance of nitrogen.

As first shown by \citet{ssw11}, broad-band colours are quite sensitive to the
abundances of C and N, particularly at shorter wavelengths.  Our calculations,
which considered the range in [Fe/H] from $-2.5$ to $-0.5$, while Sbordone et
al.~examined models for just a single metallicity, similarly predict that
increased N abundances will result in redder $U-B$ or $M_{F336}-M_{F438}$
colours --- especially along the giant branch, but also along the MS if the
metallicity is sufficiently high; see, e.g., Fig.~\ref{fig:f10}.  The same
figure shows that enhanced C abundances will cause redder $M_{F438W}-M_{F606W}$
colours and bluer $M_{F336W}-M_{F438W}$ colours.  As our exploratory study was
limited to just a single variation of the C:N and O:N abundance ratios, though
for two values of [CNO/Fe}, it would undoubtedly be worthwhile to map out the
dependence of broad-band colours along the observed C--N and O--N
anticorrelations in much finer detail.  Nevertheless, despite the limitations of
our models, they appear to be reasonably successful in explaining the
distributions of CN-weak and CN-strong stars in GCs.  This is demonstrated in
Paper II, which also presents some evidence in support of the possibility that
these systems contain C-enhanced stellar populations ([C/Fe] $\gta 0.5$).

\section*{acknowledgements}
We are indebted to Kjell Eriksson for the calculations that he carried out
to examine the dependence of synthetic spectra on the assumed model atmosphere
structures and for valuable discussions on this issue.  We also thank Poul Erik 
Nissen and Anish Amarsi for their careful readings of our paper, which resulted
in a number of improvements, and Pavel Denisenkov, John Norris, and David Yong
for helpful comments and/or useful references to published work.  LC is the
recipient of the ARC Future Fellowship FT160100402.

\section*{Data Availability}
Stellar evolutionary grids that are the basis of the isochrones that appear in
Figs.~\ref{fig:f4}--\ref{fig:f9}, along with the computer codes (in FORTRAN)
that were used to generate the isochrones on both the theoretical and observed
planes are contained in the file {\tt vecf.zip} which may be downloaded
from https://www.canfar.net/storage/list/VRmodels.  Codes are also provided to
evaluate (i) the effects on BCs of varying the microturbulent velocity and the
helium mass-fraction abundance, $Y$, and (ii) the differences in the BCs for
different mixtures of the metals relative to those for any user-selected
reference mix at the grid values of $\log$\,$g$, $\teff$, and [Fe/H].  A 
description of the contents of the zip file and instructions on how to run the 
various FORTRAN codes are given in {\tt vecf\_readme}.  This is the only other
file pertaining to the current project that should be downloaded.

\bsp   % typesetting comment
\label{lastpage}

\end{document}